\documentclass[longauth]{aa} 
\usepackage{graphicx}
\usepackage{txfonts}
\usepackage[export]{adjustbox}
\begin{document}

   \title{Herschel-ATLAS and ALMA}

   \subtitle{HATLAS\,J142935.3-002836, a lensed major merger at redshift 1.027}

   \author{Hugo Messias\inst{1,2}
          \and Simon Dye\inst{3}
          \and Neil Nagar\inst{1}
          \and Gustavo Orellana\inst{1}
          \and R.~Shane Bussmann\inst{4}
          \and Jae Calanog\inst{5}
          \and Helmut Dannerbauer\inst{6}
          \and Hai Fu\inst{7}
          \and Edo Ibar\inst{8}
          \and Andrew Inohara\inst{5}
          \and R.\,J.~Ivison\inst{9,10}
          \and Mattia Negrello\inst{11}
          \and Dominik A.~Riechers\inst{7,12}
          \and Yun-Kyeong Sheen\inst{1}
          \and Simon Amber\inst{13}
          \and Mark Birkinshaw\inst{14,15}
          \and Nathan Bourne\inst{3}
          \and Dave L. Clements\inst{16}
          \and Asantha Cooray\inst{5,7}
          \and Gianfranco De Zotti\inst{11}
          \and Ricardo Demarco\inst{1}
          \and Loretta Dunne\inst{17,9}
          \and Stephen Eales\inst{18}
          \and Simone Fleuren\inst{19}
          \and Roxana E. Lupu\inst{20}
          \and Steve J. Maddox\inst{17,9}
          \and Micha{\l} J. Micha{\l}owski\inst{9}
          \and Alain Omont\inst{21}
          \and Kate Rowlands\inst{22}
          \and Dan Smith\inst{23}
          \and Matt Smith\inst{18}
          \and Elisabetta Valiante\inst{18}
          }

   \institute{
         Universidad de Concepci\'on, Barrio Universitario, Concepci\'on, Chile
         \and Centro de Astronomia e Astrof\'{i}sica da Universidade de Lisboa,  Observat\'{o}rio Astron\'{o}mico de Lisboa, Tapada da Ajuda, 1349-018 Lisbon, Portugal
              \email{hmessias@oal.ul.pt}
         \and School of Physics and Astronomy, University of Nottingham, NG7 2RD, UK
         \and Harvard-Smithsonian Center for Astrophysics, 60 Garden Street, Cambridge, MA 02138, USA
         \and Department of Physics \& Astronomy, University of California, Irvine, CA 92697, USA
         \and Universit{\"a} Wien, Institut f{\"u}r Astrophysik, T{\"u}rkenschanzstra{\ss} e 17, 1180 Wien, Austria
         \and Astronomy Department, California Institute of Technology, MC 249-17, 1200 East California Boulevard, Pasadena, CA 91125, USA
         \and Pontificia Universidad Cat\'olica de Chile, Departamento de Astronom\'ia y Astrof\'isica, Vicu\~na Mackenna 4860, Casilla 306, Santiago 22, Chile
         \and Institute for Astronomy, University of Edinburgh, Royal Observatory, Blackford Hill, Edinburgh EH9 3HJ, UK
         \and European Southern Observatory, Karl Schwarzchild Strasse 2, Garching, Germany
         \and INAF, Osservatorio Astronomico di Padova, Vicolo Osservatorio 5, I-35122 Padova, Italy
         \and Department of Astronomy, Cornell University, Ithaca, NY 14853, USA
         \and The Open University, P.O.\ Box 197, Milton Keynes, MK7 6BJ, United Kingdom
         \and HH Wills Physics Laboratory, University of Bristol, Tyndall Avenue, Bristol BS8 1TL, UK
         \and Harvard-Smithsonian Center for Astrophysics, 60 Garden Street,Cambridge, MA 02138
         \and Astrophysics Group, Imperial College London, Blackett Laboratory, Prince Consort Road, London SW7 2AZ, UK
         \and Department of Physics and Astronomy, University of Canterbury, Private Bag 4800, Christchurch, 8140, New Zealand
         \and School of Physics and Astronomy, Cardiff University, QueensBuildings, The Parade, Cardiff CF24 3AA, UK
         \and School of Mathematical Sciences, Queen Mary, University of London, Mile End Road, London, E1 4NS, UK
         \and Department of Physics and Astronomy, University of Pennsylvania, Philadelphia, PA 19104, USA
         \and Institut d'Astrophysique de Paris, UMR 7095, CNRS, UPMC Univ. Paris 06, 98bis boulevard Arago, F-75014 Paris, France
         \and (SUPA) School of Physics \& Astronomy, University of St Andrews, North Haugh, St Andrews, KY16 9SS, UK
         \and Centre for Astrophysics Research, Science \& Technology Research Institute, University of Hertfordshire, Herts AL10 9AB, UK
             }


 
  \abstract
  {The submillimetre-bright galaxy population is believed to comprise, aside from local galaxies and radio-loud sources, intrinsically active star-forming galaxies, the brightest of which are lensed gravitationally. The latter enable studies at a level of detail beyond that usually possible by the observation facility.}
   {This work focuses on one of these lensed systems, HATLAS\,J142935.3$-$002836 (H1429$-$0028), selected in the \emph{Herschel}-ATLAS field. Gathering a rich, multi-wavelength dataset, we aim to confirm the lensing hypothesis and model the background source's morphology and dynamics, as well as to provide a full physical characterisation.}
   {Multi-wavelength high-resolution data is utilised to assess the nature of the system. A lensing-analysis algorithm which simultaneously fits different wavebands is adopted to characterise the lens. The background galaxy dynamical information is studied by reconstructing the 3-D source-plane of the ALMA CO\,(J:4$\to$3) transition. Near-IR imaging from $HST$ and $Keck$-AO allows to constrain rest-frame optical photometry independently for the foreground and background systems. Physical parameters (such as stellar and dust masses) are estimated via modelling of the spectral energy distribution taking into account source blending, foreground obscuration, and differential magnification.}
   {The system comprises a foreground edge-on disk galaxy (at $z_{\rm sp}=0.218$) with an almost complete Einstein ring around it. The background source (at $z_{\rm sp}=1.027$) is magnified by a factor of $\mu\sim8-10$ depending on wavelength. It is comprised of two components and a tens of kpc long tidal tail resembling the Antenn\ae\ merger. As a whole, the system is a massive stellar system (1.32$_{-0.41}^{+0.63}\times10^{11}\,$M$_\odot\,$) forming stars at a rate of $394\pm90$\,M$_\sun$~yr$^{-1}$, and has a significant gas reservoir ${\rm M_{ISM} = 4.6\pm1.7 \times10^{10}\,M_\sun}$. Its depletion time due to star formation alone is thus expected to be $\tau_{\rm SF}={\rm M_{ISM}/SFR}=117\pm51\,$Myr. The dynamical mass of one of the components is estimated to be $\rm{5.8\pm1.7\times10^{10}\,M_\sun}$, and, together with the photometric total mass estimate, it implies that H1429$-$0028 is a major merger system (1:2.8$_{-1.5}^{+1.8}$).}
   {}


   \maketitle
%

\section{Introduction}

Most of the sources responsible for the far-infrared (FIR) background \citep{Reach95,Puget96,Fixsen98,Lagache99} are at $z>1$ \citep{Franceschini94,Fall96,BuriganaPopa98,HauserDwek01}. Their detailed study has been limited by instrumental development: early submillimetre (submm; rest-frame FIR) studies were based on shallow and low resolution surveys \citep[e.g.][]{Scott02, Smail02, Greve04, Greve08, Magnelli09, Clements10, Dannerbauer10, Jacobs11}, but the advent of {\it Herschel} and the construction of the Atacama Large (sub-)Millimetre-Array (ALMA) are overcoming these limitations.

These recent instrumental developments have enabled systematic, detailed follow-up of the brightest of the galaxies detected in the FIR and (sub-)millimetre regime \citep[e.g.,][]{Cox11, Harris12, Karim12, Lupu12, Ivison13, Hodge13, Vieira13, Weiss13, Riechers13}, revolutionising our view of this galaxy population. Commonly referred to as submm galaxies (SMGs), they are believed to be a sporadic \citep[$\sim$100\,Myr,][]{Greve05, Tacconi06, Tacconi08} and extremely active phase of evolution \citep[star-formation rates of $\sim10^2-10^3\rm{\,M_\sun~yr^{-1}}$,][]{Ivison00, Chapman05, Coppin08, Michalowski10a, Michalowski10b, Wardlow11, Yun12, Smolcic12, Riechers13}. Whether or not this phase is responsible for the formation of the bulk of the stellar population of their descendants is still be a matter of debate \citep{Renzini06, Tacconi08, Gonzalez11}, as is the trigger for this extreme phase. Gas-rich major mergers \citep{Frayer98, Frayer99, Ivison00, Tacconi08, Daddi10b, Engel10, MenendezDelmestre13}, smooth accretion of cold gas as suggested by hydrodynamical simulations \citep{Keres05,Carilli10,Dave10,Hayward11,Hodge12} and self-regulated baryon collapse \citep{Granato04, Lapi11} have all been proposed to induce the SMG phase.

Distinguishing between, for instance, a merger event and a gas-rich clumpy disk is not trivial, as the latter may resemble a merger system in poor-resolution imaging and/or in case no velocity information is available. Hence, especially at high-redshift (when disk galaxies are believed to be clumpy), spectral/velocity and spatial detail is key \citep[e.g.,][]{ForsterSchreiber09, ForsterSchreiber11, Swinbank11, Riechers11, Ivison13}. While spectral capabilities are limited by technology, in some cases spatial resolution is boosted by nature. These cases occur when a deep gravitational potential (e.g.\ a galaxy over-density or an isolated massive galaxy) modifies the light path from a background source, inducing brightness and spatial magnification. This gravitational lensing (GL) boosts the sensitivity and resolution of our telescopes, allowing a more direct comparison with the local galaxy population \citep[see discussion in][]{Meylan06}. It is thus no surprise that GL has allowed breakthrough science in the study of distant galaxies via significantly improved detection limits and spatial resolutions \citep[e.g.,][]{Blain96, Smail02, Kneib05, SolomonBout05, Knudsen06, Tacconi06, Swinbank10}. 

Until recently, finding these rare lensed systems required deliberate searches through known galaxy over-densities where the probability of GL is higher \citep[][]{Smail97, Postman12, Furlanetto13}. However, follow-up observations of the brightest sources in under-dense regions revealed evidence of gravitational lensing \citep[e.g.,][]{Chapman02}. With the advent of wide-area (hundreds of square degrees) FIR and submm surveys, combined with powerful follow-up facilities, many such GL cases have been confirmed. This has led to simple criteria allowing efficient GL selection. Based on a small subset of bright galaxies found in the \emph{Herschel}-Astrophysical TeraHertz Large Area Survey \citep[\emph{H}-ATLAS, $\sim$570\,deg$^2$,][]{Eales10}, \citet{Negrello10} showed that a simple flux cut at 500\,$\mu$m ($S_{\rm 500\,\mu m}>100\,$mJy), followed by optical/near-IR/radio imaging analysis to discard local and radio-bright sources, is a highly efficient technique to select GL systems. Since then, more than 20 of these systems have been confirmed in \emph{Herschel} surveys \citep[e.g.][]{Conley11, Fu12, Bussmann12, Wardlow13, George13}. In parallel, observations undertaken at 1.4\,mm on the South Pole Telescope have provided a large population of GL systems \citep{Vieira13, Weiss13}.

The size of the GL sample now allows a systematic investigation of the properties of the lenses and background objects \citep[e.g.,][]{Ivison10, Frayer11, Vieira13, Weiss13, George14}, allowing direct comparison with similarly luminous local galaxies. In this work, we have obtained Atacama Large Millimetre Array (ALMA) observations of one of the lensed sources found in the \emph{H}-ATLAS, H1429$-$0028, as part of this continued effort to increase the number of GL systems with high spatial-resolution molecular data, which is still relatively scarce. With its improved detection, spectral and resolving capabilities, ALMA enables a fast and detailed view of the gas and dust in distant lensed sources, not only spatially, but also spectroscopically \citep{Vieira13,Weiss13}. This pilot study, combining {\it Herschel} and ALMA with GL, illustrates the promise of this fusion to unravel the physical processes that dominate the distant submm Universe.

The work is organised as follows: Sec.~\ref{sec:obsv} describes the source selection and the plethora of data supporting this work; Sec.~\ref{sec:results} presents the results directly obtained from the data described in the previous section; in Sec.~\ref{sec:discussion} the lensing analysis is presented along with the physical properties of both fore and background systems; Sec.~\ref{sec:conclusions} finishes with the main conclusions from this work. Throughout this work we adopt the following $\Lambda$CDM cosmology: H$_{0} = 70$\,km\,s$^{-1}$\,Mpc$^{-1}$, $\Omega_{M} = 0.3$, $\Omega_{\Lambda} = 0.7$, and a Chabrier initial mass function.

\section{Source selection and Observations} \label{sec:obsv}

\subsection{Source Selection}

HATLAS\,J142935.3$-$002836, alias H1429$-$0028, the focus of this study, was identified in the H-ATLAS coverage of the GAMA 15-hr field. With a submm flux of $S_{500\,\mu{\rm m}}=227\pm8$\,mJy, it is considerably brighter than the flux cut proposed by \citet[$S_{500\,\mu{\rm m}}>100$\,mJy]{Negrello10} to select candidate sources for gravitationally lensed systems. This source, in particular, was found to be a 160\,$\mu$m-peaker, suggesting $z\sim1$. The SPIRE data reduction is described in \citet{Pascale11}, while source extraction and flux density estimation are described in \citet{Rigby11}.

\subsection{Optical spectroscopy} \label{sec:optsp}

Long-slit spectroscopic observations at the Gemini-South telescope were taken using the Gemini Multi-Object Spectrograph-South (GMOS-S) instrument on the night of 2012 February 25 as part of program GS-2012A-Q-52 (P.I. R.~S.~Bussmann). Four observations of 15\,min each were made through a 1$\arcsec$ slit with the B600 grating. Dithering was used in both the wavelength direction and along the slit to minimise the effects of bad columns and gaps between the GMOS-S chips. The central wavelengths for the two observations were 520 and 525\,nm, and flat-field observations were interspersed between the observations at each wavelength setting. Wavelength calibration was achieved using CuAr arc lamp exposures, taken using the same instrumental set-up as for the science exposures. This provided a spectral resolution of $\approx$0.62\AA. A position angle of 70$^\circ$ East of North was used, and the detector was binned by 4 pixels in both the spectral and spatial directions.

We processed the data using standard {\sc IRAF} GMOS-S reduction routines. Since the primary aim of these observations was to obtain a spectroscopic redshift, flux calibration was not performed. We used the \textit{xcsao} routine in {\sc IRAF} to measure the spectroscopic redshift.

\subsection{Hubble Space Telescope F110W} \label{sec:hst}

A SNAPshot observation was obtained with the \emph{Hubble Space Telescope}\footnote{Based on observations made with the NASA/ESA Hubble Space Telescope, obtained at the Space Telescope Science Institute, which is operated by the Association of Universities for Research in Astronomy, Inc., under NASA contract NAS 5-26555. These observations are associated with program 12488.} (HST) on 2011 December 28$^{\rm th}$, as part of Cycle-19 proposal 12488 (P.I. Negrello), using Wide-Field Camera 3 (WFC3) with its wide $J$ filter, $F110W$. The total exposure time was 252\,s. Data were processed using the PyRAF Astrodrizzle package. Individual frames were corrected for distortion, cleaned of cosmic rays and other artifacts and median combined. The resulting $\sim2'\times2'$ image was re-sampled to a finer pixel scale of 0.0642\arcsec. The {\sc fwhm} is 0.17\arcsec\ as measured from a stellar source in the observed field.

\subsection{Keck Telescope Adaptive Optics $H$ and $K_{\rm s}$} \label{sec:keck}

We obtained sub-arcsec resolution images of H1429$-$0028 in the $H$ and $K_{\rm s}$ bands with the Keck-II laser-guide-star adaptive-optics system (LGSAO; Wizinowich et al.\ 2006). The observations took place on 2012 Feb 4 UT with the NIRC2 wide camera (0.04\arcsec\,pixel$^{-1}$) under excellent conditions (program ID: U034N2L; P.I. A.~Cooray). The only suitable tip-tilt star had $R = 15.2$ and lay 78\arcsec\ north-east of H1429$-$0028. In order to fit the star within the vignetted field for the tip-tilt sensor, we had to rotate the camera to a PA of 259.6 deg and offset H1429$-$0028 from the centre to the top-right part of the detector. We obtained 15 useful 80-s exposures in $K_{\rm s}$ and 10 useful 120-s exposures in $H$.

We used custom IDL scripts to reduce the images, following standard procedures. Briefly, after bad pixel masking, background subtraction, and flat-fielding, sky background and object masks were updated iteratively. For each frame, after subtracting a scaled median sky, the residual background was removed with 2-dimensional B-spline models. In the last iteration, we discarded frames of the poorest image quality and corrected the camera distortion using the on-sky distortion solution from observations of the globular cluster M\,92\footnote{http://www2.keck.hawaii.edu/inst/nirc2/dewarp.html}. The resolution of the final image is 0.11\arcsec\ and 0.13\arcsec\ in {\sc fwhm} for the $H$ and $K_{\rm s}$ images, respectively, as measured from two stellar sources $<21$\arcsec\ from H1429$-$0028. The two sources were nevertheless faint, and the PSF was approximated by a Gaussian with the referred widths.

\subsection{Spitzer IRAC 3.6 and 4.5$\,\mu$m} \label{sec:irac}

3.6- and 4.5-$\mu$m images were acquired using the Infrared Array Camera \citep[IRAC,][]{Fazio04} aboard \emph{Spitzer} \citep{Werner04} on 2012 September 27$^{\rm th}$ as part of the Cycle 8 GO program 80156 (P.I. A.~Cooray). The imaging involved a 38-position dither pattern, with a total exposure time of just over 1\,ks, reaching r.m.s. depths of 3.3 and 3.6\,$\mu$Jy at 3.6 and 4.5\,$\mu$m, respectively. Corrected basic calibrated data, pre-processed by the \emph{Spitzer} Science Center, were spatially aligned and combined into mosaics with a re-sampled pixel size of 0.6\arcsec\ and angular resolution of 2--2.5\arcsec, using version 18.5.0 of MOPEX \citep{MakovozMarleau05}. These data were then used for photometric measurements.

\subsection{Jansky Very Large Array 7\,GHz} \label{sec:jvla}

High-resolution 7-GHz continuum data were acquired using National Radio Astronomy Observatory’s Janksy Very Large Array\footnote{This work is based on observations carried out with the JVLA. The NRAO is a facility of the NSF operated under cooperative agreement by Associated Universities, Inc.} (JVLA) during 2011 June (proposal 11A-182; P.I. R.~J.~Ivison), in A configuration, with $64\times 2$-MHz channels in each of two intermediate frequencies (IFs), each IF with dual polarisation, recording data every 1\,s. 1505+0306 was observed every few minutes to determine complex gain solutions and bandpass corrections; 3C\,286 was used to set the absolute flux density scale. Using natural weighting, the resulting map has a $0.4'' \times 0.3''$ {\sc fwhm} synthesised beam and an r.m.s.\ noise level of 10\,$\mu$Jy\,beam$^{-1}$. 

\subsection{Z-SPEC on APEX} \label{sec:zapex}

H1429$-$0028 was observed with Z-SPEC mounted on the 12-m APEX telescope on 2--5 January 2011 as part of the {\it H}-ATLAS and Universidad de Concepci\'on collaboration (projects C-087.F-0015B-2011, P.I. G.~Orellana, and projects 087.A-0820 and 088.A-1004, P.I. R.~J.~Ivison), during the APEX P87 and P88 periods under excellent weather conditions ({\sc pwv}\,$\sim$\,0.6, ranging from 0.2 to 0.9).

Antenna pointing calibrations were performed a few times each night using a bright planet or quasar near the target producing typically $\la$\,4" corrections. Telescope focus was adjusted once each day, just after sunset, and little variation was seen throughout the observing run. To remove atmospheric fluctuations, we used a wobbler at 1.8\,Hz to switch the signal to a blank field 45\arcsec away. Data were taken in chunks of 20\,s.

Taking into account gain factors, the signal from each spectral channel was flux calibrated using observations of Uranus. This is done by building a model of the flux conversion factor (from instrument Volts to Jy) as a function of each detector's mean operating (DC) voltage (Bradford et al.\ 2009). Second-order pixel-to-pixel spectral variations ($\la5$\%) were corrected using a compilation of observations of flat-spectrum sources (J1337$-$130 and J1229+021 in this case). The spectra are considerably noisier at the lowest frequencies due to the pressure-broadening of a water line at 183\,GHz.

All errors are propagated to the source calibration using a customised pipeline developed to reduce Z-SPEC data while mounted at the Caltech Submillimetre Observatory (e.g.\ Bradford et al.\ 2009; Scott et al.\ 2011; Lupu et al.\ 2012; Zemcov et al.\ 2012).

The total integration time on source was 8.1\,hr, reaching a sensitivity of 0.8\,Jy\,s$^{1/2}$ at the bandwidth centre. The r.m.s.\ uncertainty on the final co-added spectrum ranges from 5 to 10\,mJy.

\subsection{CARMA} \label{sec:carma}

We used CARMA \citep{Bock06} to observe the CO($J$=2$\to$1) transition ($\nu_{\rm rest}$=230.5380\,GHz) toward H1429$-$0028 (proposal CX322, P.I. D.~Riechers). Based on the APEX/Z-Spec redshift of $z_{\rm spec}=1.026$, observations were made using the 3\,mm receivers with the CARMA spectral line correlator set to an effective bandwidth of 3.7\,GHz per sideband (IF range:\ 1.2--4.9\,GHz) at 5.208\,MHz (6.8\,km/s) spectral resolution, placing the redshifted CO($J$=2$\to$1) line at an IF frequency of 3.6\,GHz in the upper sideband. Observations were carried out under good 3-mm weather conditions on 2011 January 16 in a hybrid configuration between the B and E arrays (only data from 9 antennas on short baselines are used), yielding an on-source (total) observing time of 2.9\,hr (4.4\,hr; equivalent to 1.0\,hr on source with the full 15\,antenna array). The nearby quasar J1512$-$090 was observed every 20\,min for complex gain calibration. Pointing was performed at least every 2--4\,hr on nearby stars and radio quasars, using both optical and radio modes. The bandpass shape and absolute flux density scale were derived from observations of the bright quasar, 3C\,273. The resulting calibration is expected to be accurate to $\sim$15\%.

The MIRIAD package was used for data processing and analysis. The calibrated data were imaged using natural weighting, resulting in a synthesised beam of 7.1\arcsec$\times$6.1\arcsec, and an r.m.s.\ noise of 2.6\,mJy\,beam$^{-1}$ over 281.25\,MHz (365.7\,km/s).

\subsection{MAMBO-2 on IRAM-30m}

We measured the 1.2~mm continuum flux density of H1429$-$0028 using the 117-channel bolometer array, MAMBO-2, at the IRAM 30~m telescope (proposal 280-10, P.I. H.~Dannerbauer). In January 2011, the target was observed twice, each time for 8\,min, in the photometric mode of MAMBO-2. This observing mode, so-called ``on-off'', is based on the chop-nod technique where the target is placed on a reference bolometer element (on-target channel). With a beam size of $\approx$11\arcsec\ at 1.2\,mm, the continuum emission is accurately measured given the much smaller size of the source (Section~\ref{sec:mwlmorph}). Standard calibration techniques and sources --- including pointing, focus and flux calibration --- were used. Data were processed with MOPSIC, an upgrade of the MOPSI software package \citep{Zylka98}.

\subsection{ALMA}

H1429$-$0028 was observed by ALMA as part of the project 2011.0.00476.S (ALMA Cycle 0; P.I. G.~Orellana). Data from two of the five approved `science goals' --- Band 3 (centred at 107\,GHz) and Band 6 (234\,GHz) --- were observed during shared-risk time (Cycle~0) and form the core of this publication. The remaining three science goals in our proposal were not observed successfully by the end of Cycle~0.

All spectral windows (four in each set-up) were set in frequency division mode (FDM) with a 1.875-GHz bandwidth (0.488\,MHz channel width), equivalent to $\sim$2400\,km\,s$^{-1}$ ($\sim$0.6\,km\,s$^{-1}$) and $\sim$5350\,km\,s$^{-1}$ ($\sim$1.4\,km\,s$^{-1}$) in Bands 3 and 6, respectively. The tuning was based on the CARMA redshift estimate of $z=1.0271$ (Section~\ref{sec:carmares}). The total on-source integration time was about 30\,min in each band. Titan was used as a flux calibrator and J1256$-$057 (3C\,279) as a bandpass calibrator. The phase calibrator, J1408$-$078, was observed every $\sim$14\,min in Band 3 and every $\sim$12\,min in Band 6.

Data processing was done using CASA\footnote{http://casa.nrao.edu}. Initial calibration, including water vapour radiometer (WVR) corrections, phase and amplitude calibrations, were performed by the ALMA science operations team during quality assurance (QA). Our team checked the phase and amplitude steps of the calibration, and re-processed the data taking into account the new Butler-JPL-Horizons 2012 models.

\subsubsection{Band 3}

Of the two approved Band-3 science goals, only one was observed. In this science goal, the first spectral window was centred at 113.7341\,GHz to cover the $^{12}$CO\,(J:2$\to$1) transition (rest-frame 230.538\,GHz). The remaining three spectral windows were tuned to trace continuum emission at 100.879\,GHz, 102.121\,GHz, and 112.235\,GHz (rest-frame 204.482, 207.000, and 227.500\,GHz, respectively). 

This science goal was executed twice. The first execution was on 2012 May 9 with 16 antennas in the Cycle-0 `extended configuration'. Two of the 16 antennas presented visible spikes in their bandpass and the data from these antennas were deleted. To conform with the Cycle-0 specifications on the minimum number of antennas, a second execution of the science goal was made on 2012 July 28. Here, the 25 antennas were in an improved Cycle-0 `extended configuration', with baselines between $\sim$20\,m and $\sim$450\,m. Data from one antenna, DV02, was flagged by the ALMA science operations team; antenna DV08 presented a large amplitude scatter and its data were therefore also flagged.

The six observations of the phase calibrator reveal clean phase solutions with minimal phase variations ($<$8\degr\ over 14\,min) for all antennas. In two spectral windows tracing the continuum, the bandpass calibrator presented line features, necessitating the deletion of these channels. The final combined $uv$ data-set, based on the two observation runs, allows the source to be imaged at a resolution of $\sim$1.88\arcsec$\times$1.25\arcsec (natural weighting) or $\sim$1.57\arcsec$\times$1.12\arcsec (uniform weighting). The source, with a maximum extension of $\sim2\arcsec$, fits well within the primary beam of the ALMA 12\,meter antennas at this frequency ($\sim$58\arcsec), so no relative flux corrections are required across the field.

\subsubsection{Band 6}

The Band-6 science goal was executed on 2012 April 23, with 17 ALMA antennas in the Cycle 0 `extended configuration' (maximum baseline of $\sim$400~m). The adopted setup included four spectral windows: SPW0, with central frequency 242.802\,GHz, was centred on C\,{\sc i}\,$\rm{^3P_1}$~$\to$~$\rm{^3P_0}$ (rest frequency 492.161\,GHz); SPW1, with central frequency, 241.614\,GHz, was centred on CS\,(J:10$\to$9, rest frequency 489.751\,GHz); SPW2, with central frequency 227.450\,GHz, was centred on CO\,(J:4$\to$3), rest frequency, 461.041\,GHz), previously detected by APEX/Z-SPEC; SPW3 was centred at 225.950\,GHz in order to trace the source continuum. The choice of SPW tuning involved the line of interest, even though this meant some overlap of the SPWs and thus some loss of sensitivity for continuum images.

At these frequencies, Titan is clearly resolved by the longest baselines, hence the flux of the phase calibrators was determined in a subset of short-baseline antennas. Titan showed line emission in one spectral window, and the affected channels were flagged. Channels affected by atmospheric emission were also flagged. The six observations of the phase calibrator revealed clean phase solutions with minimal phase variations ($<$8\degr\ over 10\,min) for all antennas. All SPW3 data from one antenna, DV05, were flagged.

The final $uv$ data-set, based on 16 antennas with a maximum baseline of 400\,m, allowed the source to be imaged at a resolution of $\sim$0.81\arcsec$\times$0.58\arcsec (natural weighting) or $\sim$0.63\arcsec$\times$0.54\arcsec (uniform weighting). Again, no relative pointing flux corrections are required across the field. One self-calibration run was done using the CO\,(J:4$\to$3) map in order to further correct phase-delays on this dataset, improving the image quality\footnote{Further runs finding phase or amplitude solutions did not yield significant improvement.}.

\begin{figure*}
\centering
\includegraphics[width=\hsize]{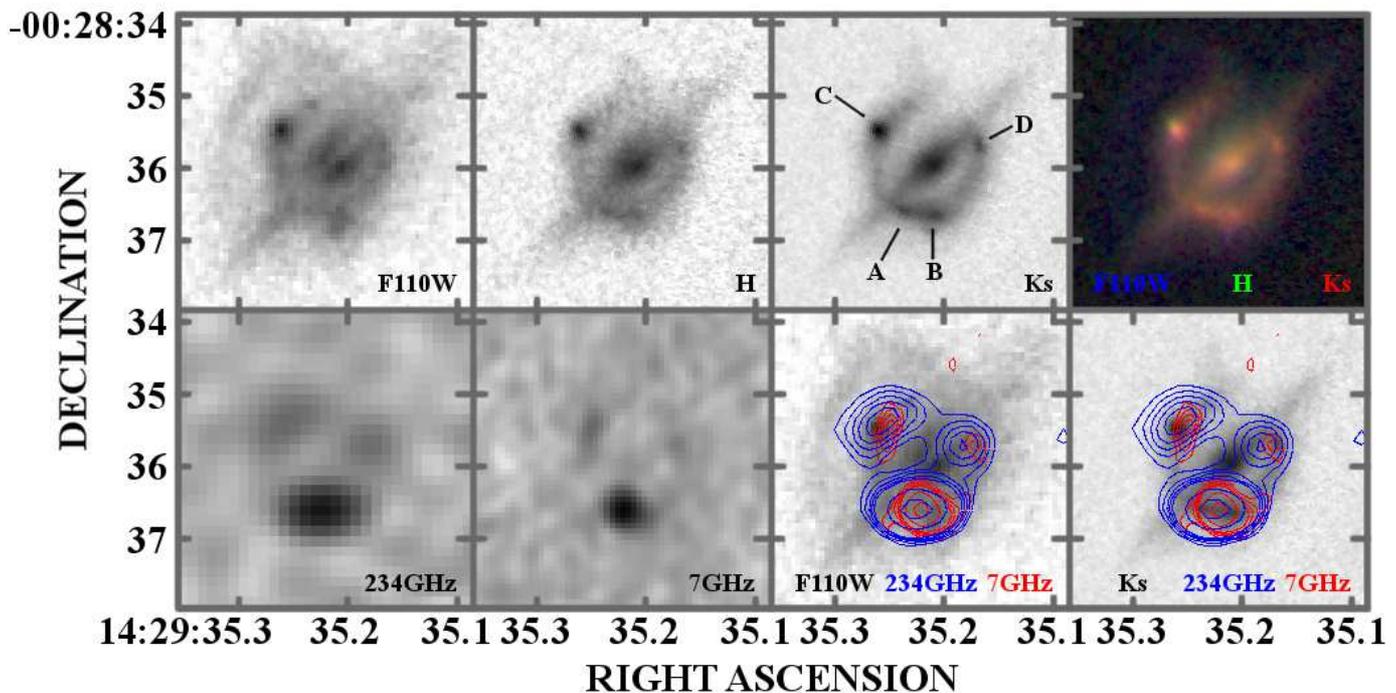}
\caption{Multi-wavelength morphology of the H1429$-$0028 system. Images are 4\arcsec\ in size. North is up; East is left. Both the foreground galaxy and the lensed galaxy --- in the form of an Einstein-ring --- are detected and resolved in the near-IR high-resolution imaging (top-row; $F110W$, $H$, and $K_{\rm s}$ are displayed with an \emph{asinh} scale, but with different flux ranges). The knot nomenclature adopted throughout the paper is indicated in the $K_{\rm s}$ imaging. Top-right panel shows a near-IR colour image (displayed with a \emph{sqrt} scale; images combined with the same flux range). Two bottom-left panels show ALMA 234- and JVLA 7-GHz continuum maps. These are compared against $F110W$ and $K_{\rm s}$ morphologies in the two bottom-right panels. Blue contours (at 3$\sigma$, $\sqrt{2}$-increments up to 675~$\mu$Jy\,beam$^{-1}$, 15$\sigma$, and 25$\sigma$, with $\sigma=78\,\mu$Jy\,beam$^{-1}$) refer to the 234~GHz continuum, while red contours (at 3$\sigma$, $\sqrt{2}$-increments up to 72~$\mu$Jy\,beam$^{-1}$, 15$\sigma$, and 30$\sigma$, with $\sigma=10\,\mu$Jy\,beam$^{-1}$) refer to the 7~GHz continuum.}
\label{fig:optnirCO}
\end{figure*}

\subsection{Data from wide-field surveys}

Given the wealth of deep wide-area surveys available today, more multi-wavelength photometry information were found in the following surveys: Sloan Digital Sky Survey \citep[SDSS,][]{York00}, VISTA Kilo-Degree Infrared Galaxy survey \citep[VIKING,][]{Sutherland12}, Wide-Field Infrared Survey Explorer \citep[WISE,][]{Wright10}, and \emph{H}-ATLAS \citep{Eales10}. From these surveys we obtained $ugriz$ (SDSS), $ZYHJK_{\rm s}$ (VIKING), 3.4--22\,$\mu$m (WISE), and 100--500\,$\mu$m (\emph{H}-ATLAS) photometry. We discuss the flux density measurements obtained from these datasets in Section~\ref{sec:mwlph}.

\section{Results} \label{sec:results}

\subsection{Multi-wavelength morphology} \label{sec:mwlmorph}

In general, seeing-limited ground-based observations of H1429$-$0028 reveal an almost point-like source. Resolving the system requires space-based, adaptive-optics- (AO-) assisted, or interferometric observations (Section~\ref{sec:obsv}). Figure~\ref{fig:optnirCO} shows a colour image of the system, made using \emph{HST}-$F110W$ (blue), Keck-AO $H$ (green) and $K_{\rm s}$ (red) imaging. These data clearly reveal a foreground source with a bulge$+$disk morphology, and an almost complete Einstein ring with a diameter of $\sim1.4\arcsec$. We identify four possible knots: two in the southern region (knots A and B); one in the north-east (knot C) and one in the north-west region (knot D).

The \emph{HST} imaging shows an additional faint arc-like feature extending from north to east, $\sim1\arcsec$ from the centre of the ring. The \emph{HST} $F110W$ filter covers the 460--678\,nm rest-frame spectral range of the background source ($z_{\rm spec}=1.027$, Section~\ref{sec:zspres}), which includes potentially bright emission lines like Halpha, so the arc could trace an extended region of star formation.

The JVLA observation of 7-GHz continuum and the ALMA observation of CO(4$\to$3) and 234-GHz continuum also provide resolved imaging of the system (see the two bottom left-hand panels in Figure~\ref{fig:optnirCO}). The JVLA and ALMA continuum maps are overlaid as contours on the \emph{HST} $F110W$ and the Keck AO $K_{\rm s}$ images (two bottom right-hand panels in Figure~\ref{fig:optnirCO}). The morphologies closely match each other, with A and B knots appearing as one. The slight offset on knot D may result from centroid position uncertainty due to the low signal-to-noise detection, but it can also be real since different components are being traced in each data-set.

Although the morphology at rest-frame optical may hint at a quad-lens system, this is only observed in the $K_s$-band. Restoring the JVLA map with Briggs weighting (robust=0), yielded a beam size of 0.29\arcsec$\times$0.26\arcsec\ and does not confirm the quad-lens morphology\footnote{Further increasing the resolution yields too noisy a map to draw any conclusion.}. Also, the relative brightness of the knots are not consistent with a simple quad-lens system, nor is there a relative-knot-flux match between the optical and mm-to-cm spectral regimes. This hints at an extended background source or a multiple-source background system, or effects induced by the foreground system. This will be discussed in Section~\ref{sec:lens}.

\subsection{Optical spectroscopy}

\begin{figure}
\centering
\includegraphics[width=\hsize]{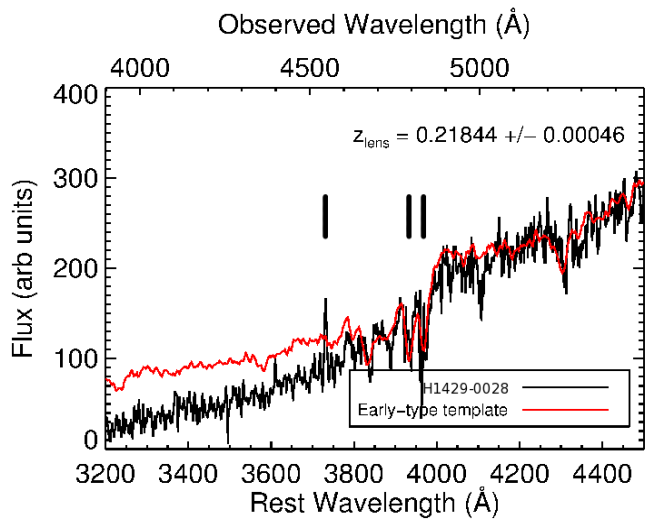}
\caption{The foreground spectrum observed with GMOS-S at Gemini-South. The black line shows the observed spectrum, while the overlaid red line shows the best-fit early-type template. The three vertical lines indicate the wavelengths of the O\,{\sc ii} and Ca H and K lines.}
\label{fig:gemini}
\end{figure}

The Gemini-South optical spectroscopy (Section~\ref{sec:optsp} and Fig.~\ref{fig:gemini}) shows the clear detection of the Ca H and K absorption lines and an O\,{\sc ii} emission line. For a template, we used a 5-Gyr-old simple stellar population from \citet{Bruzual03} with solar metallicity. While this template does not perfectly match the lensing galaxy spectra, it is sufficient to determine a precise redshift. The corresponding spectroscopic redshift of the lensing galaxy is $z=0.21844\pm0.00046$ based on the Ca absorption lines.

\subsection{Z-SPEC on APEX} \label{sec:zspres}

\begin{figure}
\centering
\includegraphics[width=\hsize]{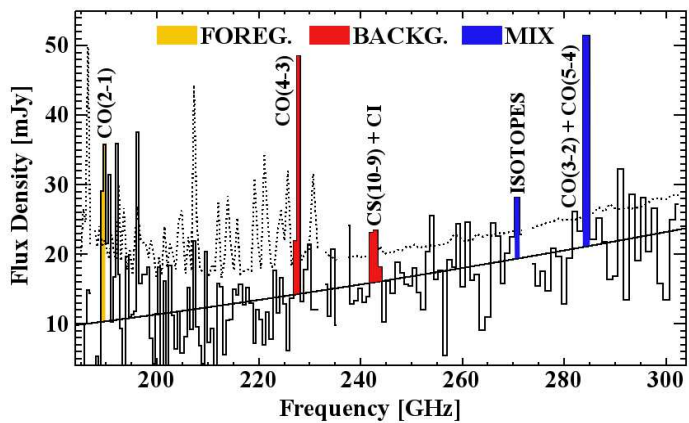}
\caption{The Z-Spec spectrum of H1429-0028 with observed frequency on the x-axis and flux density (mJy) on the y-axis. Reliably detected lines include CO\,(J:4$\to$3) and CO\,(J:5$\to$4) at $z=1.026\pm0.002$. The channel where CO\,(J:5$\to$4) from the background source is observed is likely contaminated by CO\,(J:3$\to$2) from the foreground source. Colours represent channels with line emission from the foreground source (yellow), background source (red), or a mix from both (blue, see text for more details). The solid line shows the fit to the continuum, while the dashed line represents the $1\sigma$ error above the continuum.}
\label{fig:zspec}
\end{figure}

The APEX/Z-Spec spectrum of H1429$-$0028 is shown in Figure~\ref{fig:zspec}. The two reliably-detected lines are identified as CO\,(J:4$\to$3) and CO\,(J:5$\to$4), yielding a redshift of 1.026$\pm$0.002 for the background source. The integrated fluxes are $37.6\pm8.8$\,Jy\,km\,s$^{-1}$ for CO\,(J:4$\to$3), and $40.0\pm5.9$\,Jy\,km\,s$^{-1}$ for CO\,(J:5$\to$4). The continuum was considered to be power-law ($f\propto\nu^\alpha$) with a spectral-index of $\alpha=1.76\pm0.23$.

The redshift of the foreground source implies that the CO\,(J:3$\to$2) transition falls on the same Z-SPEC channel as the background CO\,(J:5$\to$4) emission (at an expected frequency separation of $\Delta\nu\sim0.4\,$GHz). Also, the foreground CO\,(J:2$\to$1) emission line falls in a very noisy part of the spectrum (at 189\,GHz), showing a detection significance level of $\sim2.8\sigma$ ($32.9\pm11.7$\,Jy\,km\,s$^{-1}$). Attempting to constrain the flux of the foreground CO\,(J:3$\to$2) transition provides a broad flux range ($4.2\pm1.5-32.9\pm11.7$\,Jy\,km\,s$^{-1}$) assuming CO\,(J:2$\to$1) to CO\,(J:3$\to$2) ratios observed for spiral galaxies \citep{Braine93,Mao10}. Hence, we expect the background CO\,(J:5$\to$4) integrated flux to be $<35.7\pm6.1$\,Jy\,km\,s$^{-1}$, which is still consistent with CO\,(J:5$\to$4) being responsible for all the channel flux ($40.0\pm5.9$\,Jy\,km\,s$^{-1}$).

Since our ALMA Band 9 observations, targeting CO\,(J:12$\to$11) and J:11$\to$10) at $z=1.027$, were not observed, we defer any study of the CO ladder to a future analysis when more transitions have been observed.

Finally, we highlight the channels with flux levels at the $\sim2\sigma$ level corresponding to the summed contribution of multiple transitions. At $\sim$242\,GHz, the background CS\,(J:10$\to$9) and [C\,I\,$\rm{^3P_1}$~$\to$~$\rm{^3P_0}$] transitions yield together a flux of $14.1\pm5.5$\,Jy\,km\,s$^{-1}$. These two transitions were observed by this project, and further discussion is given in Section~\ref{sec:almaline}. At $\sim$242\,GHz, foreground $^{13}$CO\,(J:3$\to$2) and C$^{18}$O\,(3$\to$2) transitions couple with the background C$^{18}$O\,(5$\to$4) transition, yielding a flux of $8.8\pm4.2$\,Jy\,km\,s$^{-1}$. Also, two absorption features may be observed at $\sim$265\,GHz and $\sim$300\,GHz (at $\sim2\sigma$). These frequencies match, among others, those of foreground H$_2$O, HCO, NH$_3$, and CH$_3$OH transitions. Future observations will test the reality of these absorption features.

\subsection{CARMA} \label{sec:carmares}

The system was not spatially resolved by our CARMA observations. The CO\,(J:2$-$1) line is offset from the initial tuning, implying an improved redshift estimate (in comparison to that obtained from Z-SPEC) of $z=1.0271\pm0.0002$. The velocity-integrated line flux is $14.4\pm 1.8$\,Jy\,km\,s$^{-1}$.

\subsection{ALMA Data: Line emission} \label{sec:almaline}

\begin{figure*}
\centering
\includegraphics[width=\hsize]{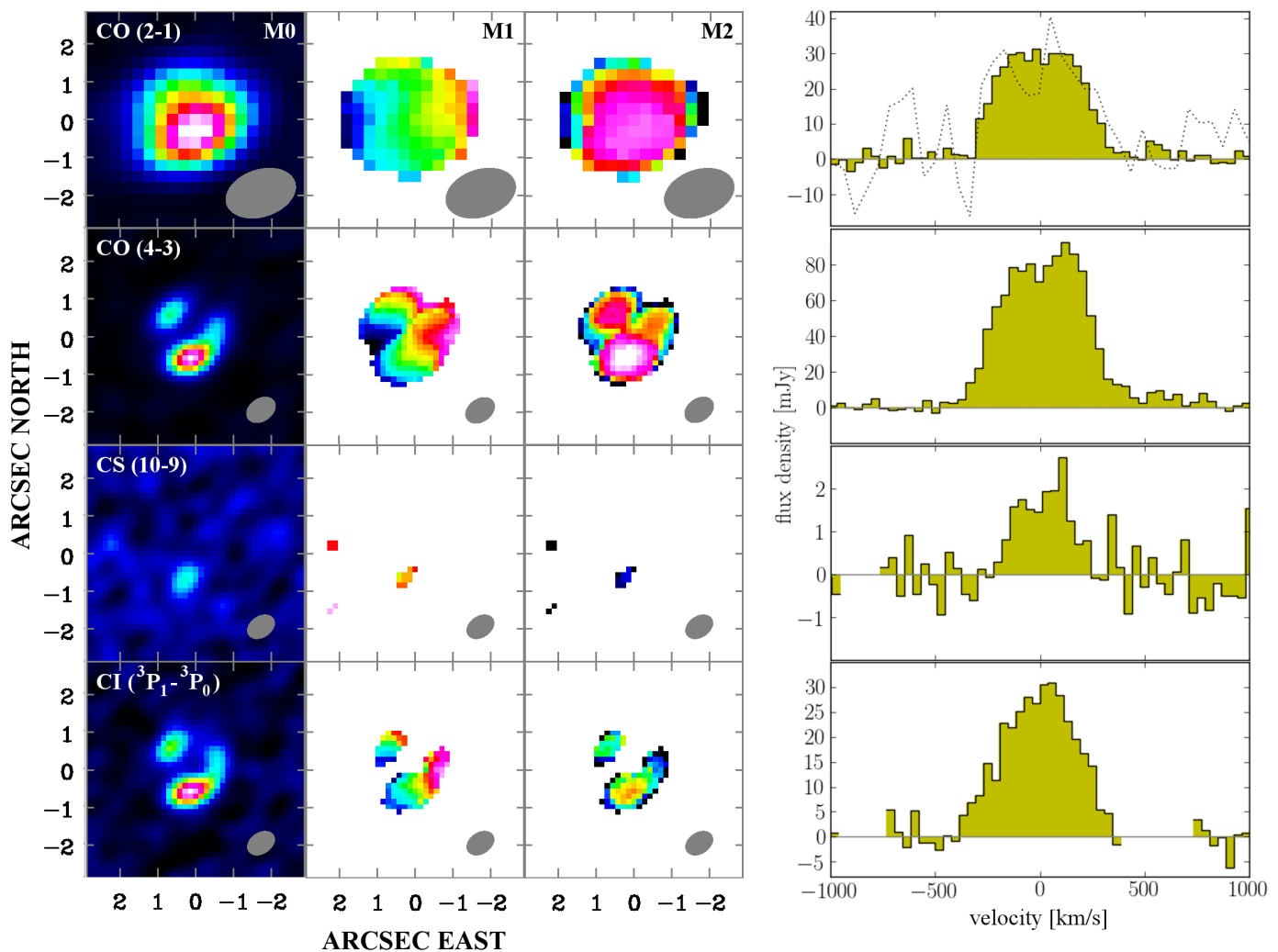}
\caption{Moment maps and line profiles of the emission lines detected in H1429$-$0028 as observed by ALMA: CO(2$\to$1) on the first (upper) row; CO(4$\to$3) on the second row; CS(10$\to$9)\ on the third row; [CI\,$\rm{^3P_1}$~$\to$~$\rm{^3P_0}$]\ on the fourth row. The columns show different image moments: moment 0 (M0, velocity-integrated flux, left), 1 (M1, velocity map, middle), and 2 (M2, dispersion map, right). Natural weighting was adopted to produce the moments. The beam is shown at the lower right in each panel as a shaded ellipse. Axes units are arcseconds. Colour scales of M0 are from -0.3 to 8 (first row), 18 (second), 2 (third), 5\,Jy\,km\,s$^{-1}$ (fourth). Colour scales in M1 and M2 are, respectively, $-$200 to 200\,km\,s$^{-1}$ and 0 to 170\,km\,s$^{-1}$. The right-most column shows the line-profiles at a spectral resolution of 40\,km\,s$^{-1}$. The first row also shows the CARMA spectrum as a dotted line.}
\label{fig:mom}
\end{figure*}

Fig.~\ref{fig:mom} shows the moment maps of the four lines observed with ALMA: CO(2$\to$1), CO(4$\to$3), [C\,I\,$\rm{^3P_1}$~$\to$~$\rm{^3P_0}$] and CS(10$\to$9). Each row shows the moment maps of a single spectral line, while each column shows different moments (left to right): integrated spectral line flux (moment 0, M0), velocity field (moment 1, M1), and velocity dispersion (moment 2, M2). The overall Einstein ring morphology is seen in the higher-frequency lines, and the emission can be separated into three main components: the brightest region in the south (knots A$+$B) is extended towards the north-west (knot D), while a third component is observed in the north-east (knot C). The CO(2$\to$1) emission is close to unresolved, but the clean component map shows the presence of the A$+$B and C knots. All emission lines are detected in the brightest southern component. The CS(10$\to$9) line, however, is not reliably detected toward the C and D knots. 

\begin{figure}
\centering
\includegraphics[width=\hsize]{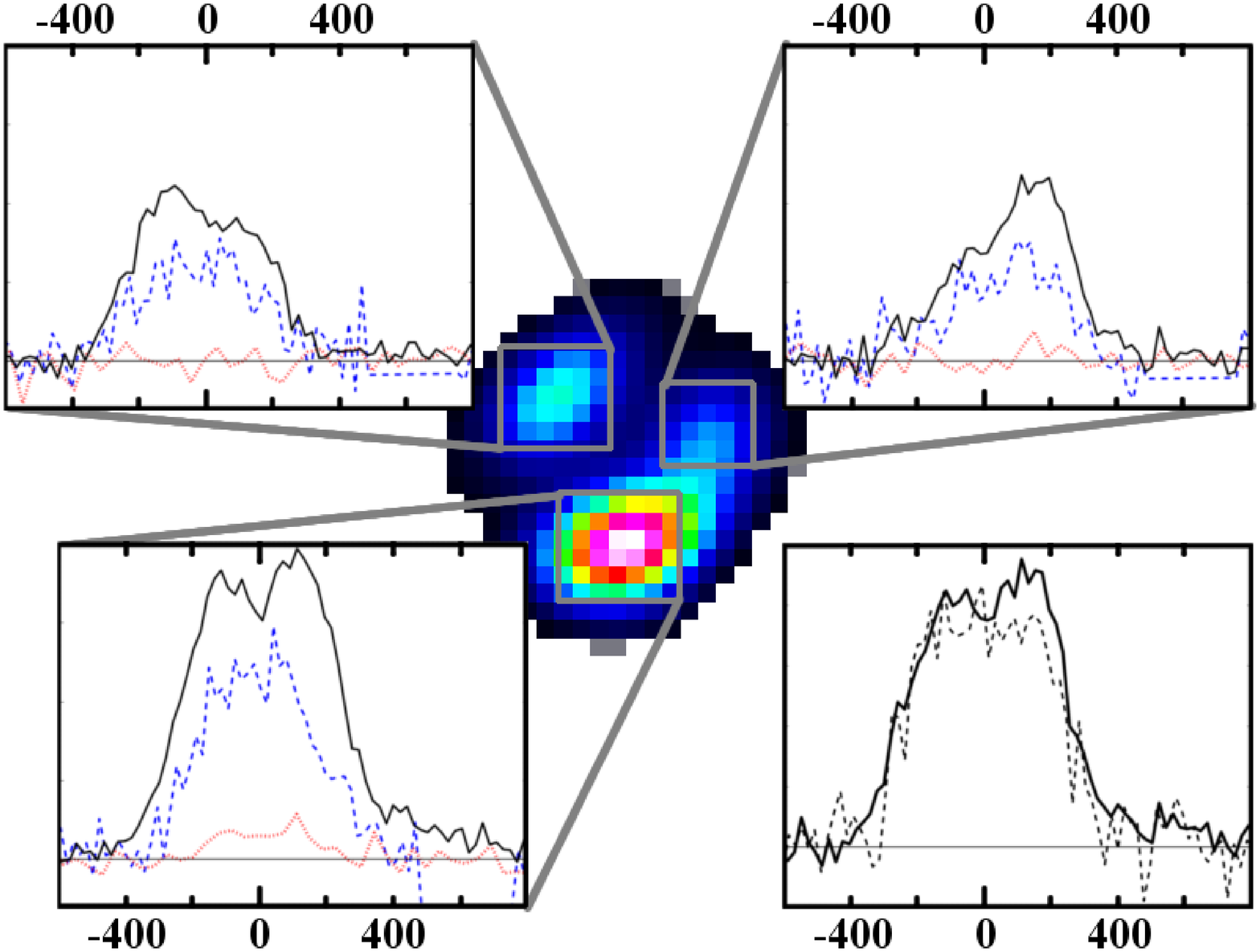}
\caption{Line profiles of the emission lines detected in H1429$-$0028 on each of the specified knots A$+$B, C and D (grey boxes): CO(4$\to$3) as solid black line; CS(10$\to$9) as a dotted red line (scaled up by a factor of 2); and [CI\,$\rm{^3P_1}$~$\to$~$\rm{^3P_0}$] as dashed blue line (scaled up by a factor of 2). The bottom right panel compares the line profiles of the overall CO(2$\to$1) (dashed line, scaled up by a factor of 2.5) and CO(4$\to$3) (solid line) emission. Different spectral resolutions are considered (20\,km\,s$^{-1}$ for CO(2$\to$1), CO(4$\to$3) and C\,{\sc i}, and 40\,km\,s$^{-1}$ for CS(10$\to$9)). The $y$-axes have the same span in all panels, except the bottom right one. The x-axis range is $-$600 to 800\,km\,s$^{-1}$ in all panels. The horizontal solid line indicates the zero-flux level. The background colour image is the CO(4$\to$3) moment-0 map from Fig.~\ref{fig:mom}.}
\label{fig:prof}
\end{figure}

The line profiles are also shown in the right-hand column in Fig.~\ref{fig:mom}. The CO shows a double-peaked or plateau profile with a redshifted tail. That is also observed individually in the A$+$B knot (Fig.~\ref{fig:prof}). Although it is difficult to claim the same for C\,I and CS, the peaks in the latter do appear to align with those of CO. The line emission from knot D is predominantly observed systematically redshifted.

Table~\ref{tab:lineflux} details the emission line parameters for the system as a whole and for each knot. Line luminosities of a transition ($L'_{\rm trans}$) are estimated as follows:
\[L'_{\rm trans}=3.25\times10^7~S_{\rm trans}~\Delta{V} ~\nu^{-2}_{\rm obs}~D^2_L~(1+z)^{-3},\]
measured in $\rm{K\,km\,s^{-1}\,pc^2}$, where the integrated flux $S_{\rm CO}\Delta{V}$ is in Jy\,km\,s$^{-1}$, the observed frequency $\nu_ {\rm obs}$ is in GHz, and the luminosity distance $D_L$ is in Mpc \citep[e.g.,][]{Solomon97}.

\begin{table*}
\caption{Observed lines in H1429$-$0028.}
\label{tab:lineflux}
\centering
\begin{tabular}{lcrrrrrr}
\hline\hline
Line & Region & obs.\ (rest) freq.\tablefootmark{a} & redshift & Line peak & Integ.~flux & Line {\sc fwhm}\tablefootmark{a} & $L'_{\rm trans}$\tablefootmark{b} \\
 &  & [GHz] &  & [mJy] & [Jy km\,s$^{-1}$] & [km\,s$^{-1}$] & [1E10~K km\,s$^{-1}$.pc$^2$] \\
\hline
CO\,(J:2$\to$1) & Total & 113.733$\pm$0.001 & 1.027011$\pm$1.8E-5 & 32.0$\pm$6.2 & 15.1$\pm$1.0 & 469$\pm$12 & 21.2$\pm$1.4\\
& A$+$B & (230.538) & & 18.1$\pm$3.4 & 7.98$\pm$0.43 & 507$\pm$16 & 11.22$\pm$0.60 \\
& C & & & 10.2$\pm$2.0 & 3.58$\pm$0.37 & 406$\pm$17 & 5.04$\pm$0.52 \\
& D & & & 4.5$\pm$2.1 & 1.61$\pm$0.35 & 461$\pm$36 & 2.26$\pm$0.49 \\
CO\,(J:4$\to$3) & Total & 227.433$\pm$0.004 & 1.027151$\pm$3.5E-5 & 89.9$\pm$9.9 & 43.9$\pm$3.7 & 481$\pm$13 & 15.5$\pm$1.3 \\
& A$+$B & (461.041) & & 46.1$\pm$4.6 & 24.4$\pm$1.9 & 496$\pm$13 & 8.60$\pm$0.65 \\
& C & & & 21.8$\pm$4.2 & 10.2$\pm$1.7 & 445$\pm$18 & 3.57$\pm$0.60 \\
& D & & & 17.5$\pm$3.8 & 6.0$\pm$1.4 & 383$\pm$19 & 2.09$\pm$0.50 \\
CS\,(J:10$\to$9) & Total & 241.568$\pm$0.027 & 1.02738$\pm$2.2E-4 & 3.0$\pm$3.2 & 0.73$\pm$0.53 & 347$\pm$83 & 0.23$\pm$0.17 \\
& A$+$B & (489.751) & & 3.2$\pm$2.6 & 0.79$\pm$0.46 & 347$\pm$83 & 0.25$\pm$0.14 \\
& C & & & $<$2.1 & -- & -- & -- \\
& D & & & $<$2.1 & -- & -- & -- \\
CI\,($\rm{^3P_1}$~$\to$~$^3\rm{P_0}$) & Total & 242.819$\pm$0.010 & 1.027181$\pm$7.2E-5 & 33.5$\pm$9.6 & 13.4$\pm$2.0 & 479$\pm$27 & 4.14$\pm$0.60 \\
& A$+$B & (492.161) & & 18.0$\pm$5.4 & 7.5$\pm$1.1 & 447$\pm$21 & 2.32$\pm$0.34 \\
& C & & & 8.9$\pm$5.3 & 3.25$\pm$0.89 & 455$\pm$36 & 1.00$\pm$0.27 \\
& D & & & 7.2$\pm$3.4 & 2.38$\pm$0.92 & 404$\pm$39 & 0.73$\pm$0.28 \\
\hline
\end{tabular}
\tablefoot{Information of the different line transitions targeted by our ALMA observations. Errors in columns 4 to 8 indicate the $1\sigma$ uncertainty. Upper limits are set at the $3\sigma$ level. Fluxes measured in the clean component map using natural weighting. \\
\tablefoottext{a}{The observed frequency as computed assuming a Gaussian profile. The value in parenthesis refers to the rest-frame line frequency.}\\
\tablefoottext{b}{The adopted redshift is $z=1.027$ ($\rm{D_L}=6828.3$\,Mpc).}
}
\end{table*}

We would like to highlight the flux density agreement between ALMA observations and those of Z-SPEC and CARMA. For instance, even the faint detection of the joint emission from C\,I\,$\rm{^3P_1}$~$\to$~$\rm{^3P_0}$ and CS(10$\to$9) in the Z-SPEC spectrum yields a flux estimate ($14.1\pm5.5$\,Jy\,km\,s$^{-1}$) in good agreement with what is estimated from the ALMA observations ($14.2\pm2.0$\,Jy\,km\,s$^{-1}$).

As mentioned before, the Z-Spec data suggest an upper limit on CO(5$\to$4) luminosity of $L'_{\rm trans}<8.04^{+1.37}$. Together with the other observed lines, this yields line ratios of $\frac{L'_{\rm CO(2\to1)}}{L'_{\rm CO(4\to3)}}=1.37\pm0.15$, $\frac{L'_{\rm CO(2\to1)}}{L'_{\rm CO(5\to4)}}>2.5_{-1.5}$, and $\frac{L'_{\rm CO(4\to3)}}{L'_{\rm CO(5-4)}}>1.77_{-0.39}$. Hence, H1429$-$0028 has values consistent with line ratios observed in SMGs and QSOs \citep[within the natural scatter of these populations,][]{CarilliWalter13}.

\subsection{ALMA: continuum emission} \label{sec:almacont}

Continuum-only images were made individually for each spectral window (in each band) after discarding channels with line emission from the transitions presented above. In addition, we then created a higher signal-to-noise continuum map by combining all line-free channels to obtain effective bandwidths of 5.0\,GHz and 4.7\,GHz in bands 3 and 6, respectively. It should be noted that two of the spectral windows were positioned in the lower side band, and the other two in the upper side band, meaning a frequency gap of 8.2 and 10.4\,GHz in bands 3 and 6, respectively.

Table~\ref{tab:cont} lists the total and knot continuum flux densities in each spectral window and each band. The total flux estimates yield a millimetre spectral index (where $f\propto\nu^{\beta}$) of $\beta=3.29\pm0.40$. 

The number of channels used in SPW0 of the band-6 observations (B6-0) is reduced due to atmospheric line flagging, resulting in a larger flux error. In band 3 SPW0 (B3-0), the r.m.s.\ level is high due to the reduced number of line-free channels in this spectral window targeting CO(2$\to$1). As a result, while computing the millimetre spectral index (where $f\propto\nu^{\beta}$) we adopt the flux density obtained for band-6 SPW1 to compare with that for band 3 SPW1. Such comparison implies a spectral index of $\beta=3.35\pm0.84$. Comparing the \emph{Herschel} 500\,$\mu$m flux (227$\pm$8\,mJy) with that at 1.28\,mm, one obtains a spectral index of $\beta=3.89\pm0.41$ (having factoring in a conservative 10\% flux calibration uncertainty for {\it Herschel}). The difference in the slopes, even though consistent within the errors, is expected to result from free-free emission contributing to the 2.8-mm continuum \citep[e.g.][]{Thomson12}.

\begin{table*}
\caption{Continuum emission from H1429$-$0028.}
\label{tab:cont}
\centering
\begin{tabular}{crrrrrr}
\hline\hline
SPW & $\lambda_{\rm cent}$ & r.m.s. & Total & A$+$B & C & D \\
 & [mm] & [mJy] & [mJy] & [mJy] & [mJy] & [mJy] \\
\hline
 B3\tablefootmark{a} & 2.80 & 0.025 & 0.54$\pm$0.11\tablefootmark{b} & 0.374$\pm$0.044 & 0.107$\pm$0.044 & $<$0.075 \\
 B3-0 & 2.64 & 0.063 & 0.72$\pm$0.22 & 0.49$\pm$0.13 & 0.20$\pm$0.16 & $<$0.19 \\
 B3-1 & 2.67 & 0.038 & 0.47$\pm$0.13 & 0.309$\pm$0.067 & 0.058$\pm$0.067 & $<$0.11 \\
 B3-2 & 2.94 & 0.070 & 0.22$\pm$0.10 & 0.22$\pm$0.10 & $<$0.21 & $<$0.21 \\
 B3-3 & 2.97 & 0.054 & 0.56$\pm$0.16 & 0.38$\pm$0.11 & 0.111$\pm$0.094 & $<$0.16 \\
 B6   & 1.28 & 0.078 & 5.86$\pm$0.99\tablefootmark{b} & 3.69$\pm$0.28 & 1.31$\pm$0.27 & 0.71$\pm$0.16 \\
 B6-0 & 1.23 & 0.217 & 7.1$\pm$1.3 & 4.12$\pm$0.86 & 1.40$\pm$0.69 & 0.79$\pm$0.47 \\
 B6-1 & 1.24 & 0.137 & 6.15$\pm$0.83 & 3.67$\pm$0.48 & 1.39$\pm$0.50 & 0.76$\pm$0.31 \\
 B6-2 & 1.32 & 0.132 & 6.12$\pm$0.74 & 3.92$\pm$0.48 & 1.17$\pm$0.42 & 0.95$\pm$0.30 \\
 B6-3 & 1.33 & 0.115 & 5.29$\pm$0.57 & 3.26$\pm$0.40 & 1.22$\pm$0.28 & 0.50$\pm$0.16 \\
\hline
\end{tabular}
\tablefoot{Information of the different continuum bands targeted by our ALMA observations. Numbers in parenthesis indicate the $1\sigma$ uncertainty. Upper limits are set at the $3\sigma$ level. Flux densities measured in the clean component map using natural weighting. \\
\tablefoottext{a}{Considering only the line-free spectral windows 1, 2 and 3.} \\
\tablefoottext{b}{The errors consider a conservative 15\% error, added in quadrature, to account for flux density calibration uncertainty, as suggested by ALMA staff.}
}
\end{table*}

\subsection{Multi-wavelength photometry} \label{sec:mwlph}

Photometry was gathered in a larger range of the spectrum (from $U$ to 4\,cm). The system is of course composed of both the foreground and background galaxies, which contribute differently in each spectral regime. A detailed study of the foreground and background SEDs contribution to the total SED is presented in Section~\ref{sec:fbsed}.

The SDSS fluxes refer to the `model magnitudes'\footnote{The SDSS magnitudes $u$ and $z$ have been converted to AB magnitudes by adding, respectively, -0.04 and 0.02. The $gri$ photometry is expected to be close to AB. \emph{in}: http://www.sdss.org/dr5/algorithms/fluxcal.html
} provided in the SDSS-DR9 \emph{Explore Home} \citep{Ahn12}. This is what is suggested by the SDSS team when the goal is to estimate galaxy colours\footnote{http://www.sdss.org/dr7/algorithms/photometry.html
} as it is done in Section~\ref{sec:fbsed}. The \emph{HST}-$F110W$ flux is that within the elliptical aperture used for the deblending analysis described in the next paragraph. The VIKING, \emph{Spitzer} IRAC, and WISE fluxes are measured within a $\sim$8\arcsec diameter aperture. The adopted aperture size does not include neighbour flux given that the closest sources are at a projected distance of $\sim$12\arcsec\ (very faint source) and $\sim$17\arcsec.

Finally, given the angular size of H1429$-$0028 being much smaller than the FWHM of the \emph{Herschel} bands, \emph{Herschel} SPIRE flux densities are those directly derived by the Multi–band Algorithm for source eXtraction (MADX, Maddox et al. in prep.), while \emph{Herschel} PACS flux densities are estimated for the SPIRE source position within apertures of 10\arcsec\ (100\,$\mu$m) and 15\arcsec\ (160\,$\mu$m). This procedure is described in detail in \citet{Rigby11}. PACS and SPIRE data reduction are described, respectively, in \citet{Ibar10} and \citet{Pascale11}.

\begin{figure*}
\centering
\includegraphics[scale=0.3]{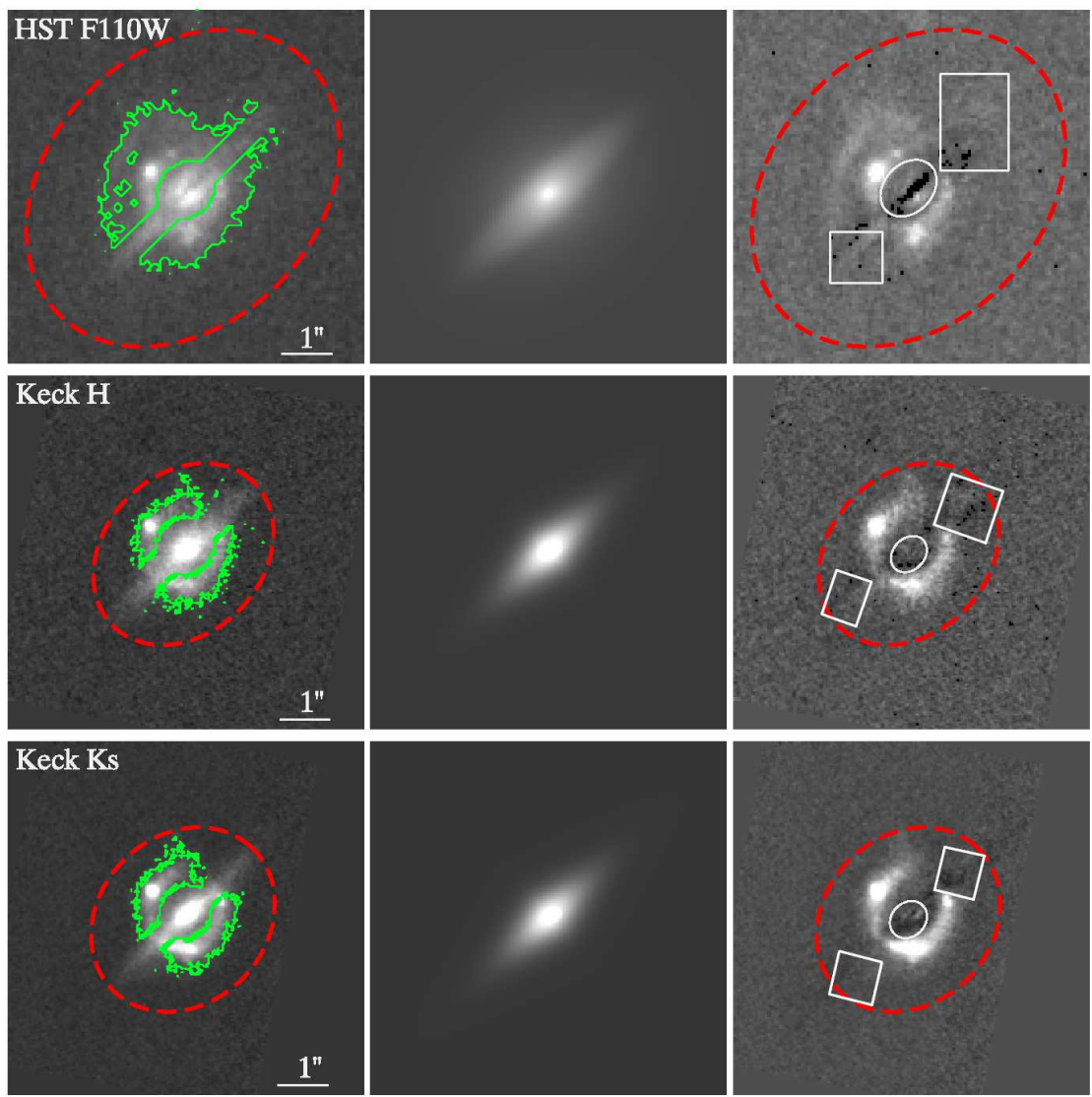}
\caption{Using {\sc galfit} to estimate the foreground emission profile. North is up; East is left. Left-hand side panels show original imaging from {\it HST} F110W (upper row), Keck-AO $H$ (middle row), and Keck-AO $K_{\rm s}$ (bottom row). The right-hand side panels show the residuals after foreground emission subtraction using the model in the middle panels. The green contours in the left panels delimit the mask used to indicate pixels masked out in the {\sc galfit} analysis. The red dashed ellipses encompass the region where the flux was estimated. The extra squared regions on the right panels flag out over-subtracted regions for improved photometry.}
\label{fig:galfit}
\end{figure*}

Judging from Fig.~\ref{fig:optnirCO}, the foreground and background sources are more similar in brightness at rest-frame optical than at long wavelengths, where the background emission dominates. In order to estimate the flux of each of the two sources in the high resolution rest-frame optical imaging (i.e.\ in the \emph{HST} $F110W$ and Keck-AO $H$ and $K_{\rm s}$ band observations), we have used {\sc galfit}\footnote{http://users.obs.carnegiescience.edu/peng/work/galfit/galfit.html} \citep[version 3.0.4,][]{Peng10} to fit and estimate foreground and background fluxes. We have masked out the image pixels dominated by background emission (green contours in left hand-side panels in Fig.~\ref{fig:galfit}). The disk galaxy light-profile is considered to be composed of an edge-on disk profile plus a S\'ersic index profile (the latter is used to fit the bulge component). Even though the residuals (right hand-side panels in Fig.~\ref{fig:galfit}) show over-subtracted regions (likely induced by the dust lane in the foreground galaxy, Fig.~\ref{fig:optnirCO}), we expect this not to be relevant to our analysis, as these regions can be masked out while estimating the background flux (solid white boxes and ellipse in right hand-side panels). Finally, comparing `original' against `foreground-removed' imaging, we estimate background-to-total flux density fraction within the same aperture (red dashed ellipse in the figure). These fractions are $20.9\pm1.3\%$ at $1.1\,\mu$m ($F110W$), $29.7\pm0.1\%$ at $1.6\,\mu$m ($H$-band), and $40.8\pm0.1\%$ at $2.2\,\mu$m ($K_{\rm s}$-band).

\section{Discussion} \label{sec:discussion}

\subsection{Lens model} \label{sec:lens}

As referred to in Section~\ref{sec:mwlmorph}, despite presenting a quad-lens-like knot positioning in $K_{\rm s}$-band, the relative brightness of the knots is troublesome. While in the rest-frame optical the C knot appears much brighter than the A$+$B knots, the opposite happens in the CO and mm-to-radio continuum emission (Fig.~\ref{fig:optnirCO} and Table~\ref{tab:lineflux}). A few relevant scenarios may explain such multi-wavelength relative knot brightnesses: (i) the background source is extended or clumpy; (ii) significant foreground obscuration is affecting the emission of knots A, B, and D at rest-frame optical wavelengths; (iii) C is being micro-lensed; (iv) a non-standard dark-matter halo structure; (v) variability.

For the current discussion, we will assume that scenarios (iii) and (iv) do not apply given the lack of data to address such possibilities, but we acknowledge their likelihood. Based on the fact that the JVLA observations in 2011 June and those of ALMA between 2012 April and July show similar morphology (A$+$B knot being the brightest), which is distinct from the $F110W$-to-$K_{\rm s}$ imaging (C knot being the brightest) taken between 2011 December and 2012 February, we can safely assume variability is not responsible for the discrepant multi-wavelength morphology. Consequently, scenarios (i) and (ii) are those addressed henceforth.

\subsubsection{Lens characterisation}

Our analysis is done with the enhanced version of the semi-linear inversion (SLI) method algorithm originally derived by \citet{Warren03} and described in \citet{Dye14}. This code does not assume any {\it a priori} background morphology and allows multiple datasets to be simultaneously reconstructed using the same lens mass model. Given the likelihood of foreground obscuration at rest-frame optical wavelengths, the images given as input are the velocity-integrated CO(4$\to$3) line map and the 7-GHz continuum map. Both maps were reconstructed with similar beam sizes and equal pixel scales.

The lens modelling we pursue assumes an elliptical power-law mass density profile \citet{KassiolaKovner93}:
\begin{equation}
\kappa = \kappa_0 ( \tilde{r} / 1\,\rm{kpc} )^{1-\alpha},\label{eq:kappa}
\end{equation}
where: $\kappa$ is the surface mass density; $\kappa_0$ is the normalisation surface mass density; $\tilde{r}$ is the elliptical radius defined by $\tilde{r} = x'^2 + y'^2/\epsilon^2$ ($\epsilon$ being the lens elongation defined as the ratio of the semi-major to semi-minor axes); and $\alpha$ is the power-law index relating the volume mass density, $\rho$, with radius, $r$: $\rho \propto r^{-\alpha}$. The profile is also described by the orientation of its semi-major axis ($\theta$, measured counter-clockwise from North of the semi-major axis) and the position of the mass center in the image-plane ($x_c$, $y_c$). External shear is not considered, because no evidence for its presence was found during the analysis.

\begin{figure}
\centering
\includegraphics[scale=0.5,angle=-90]{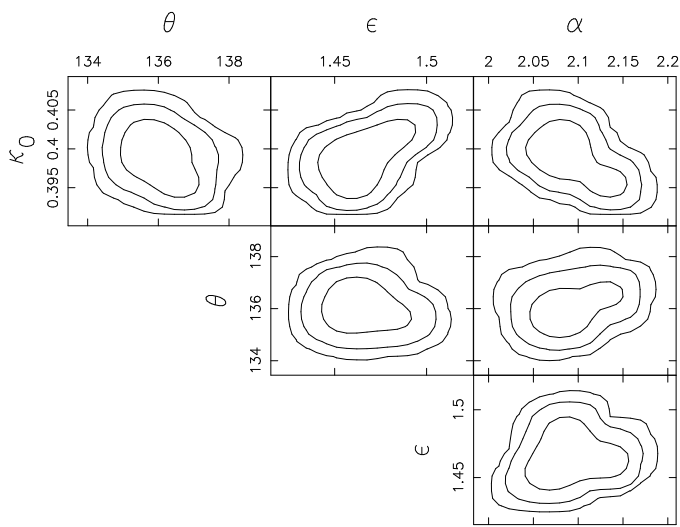}
\caption{The parameter confidence space. Parameters shown are the normalisation surface mass density ($\kappa_0$ in units of $10^{10}~$M$_\odot~$kpc$^{-2}$), the lens elongation ($\epsilon$), the mass profile power-law index ($\alpha$), and the orientation of the lens semi-major axis ($\theta$ in degrees measured counter-clockwise from North).}
\label{fig:parconf}
\end{figure}

The geometric average of the Einstein radius, $\theta_{\rm E}$, is computed as:
\[ \left(\frac{\theta_{\rm E}}{1\,{\rm kpc}}\right) = \left( \frac{2}{3-\alpha} \frac{1}{\sqrt(\epsilon)} \frac{\kappa_0}{\Sigma_{\rm CR}}\right) ^{\frac{1}{\alpha-1}}\]
where $\Sigma_{\rm CR}$ is the critical surface mass density \citep[e.g.,][]{Schneider92}. The best fit parameters resulting from the analysis referred above are $\kappa_0=0.399_{-0.006}^{+0.005}\times10^{10}~$M$_\odot~$kpc$^{-2}$, $\alpha=2.08_{-0.05}^{+0.07}$, $\epsilon=1.46_{-0.03}^{+0.04}$, $\theta=135.9_{-1.0}^{+1.2}~$deg, and $\theta_{\rm E}=2.18_{-0.27}^{+0.19}\,$kpc ($0.617_{-0.075}^{+0.054}$\arcsec). The confidence limits are shown in Fig.~\ref{fig:parconf}.


In order to assess how well each dataset has been reconstructed, we computed (i) the significance (flux density to clean-residual r.m.s.\ ratio) of the flux density in pixels at $>1\sigma$ in the masked residual image, and (ii) the number of residual image pixels with a significance greater than 2.0 as a fraction of the total number of masked image pixels. This is instead of quoting Bayesian evidence, which is meaningless when not comparing models, and is instead of quoting $\chi^2$, which, owing to covariance between the source plane pixels from regularisation in the SLI method, is subject to an ill-defined number of degrees of freedom \citep[$\nu$; see][]{Suyu06}. Regarding assessment (ii), we measure a fraction of $>2\sigma$ residuals of 0.2\% and 0.5\% for the 7\,GHz and CO\,(J:4$\to$3) data respectively. We have verified that both datasets are well described by Gaussian statistics and therefore attribute the fact that this is significantly lower than the expected fraction of 4.6\% to the fact that the SLI method fits away some of the image noise. Assessment (i) yields a residual flux significance of 1.9 and 0.4 for the 7\,Ghz and CO\,(J:4$\to$3) datasets respectively.

An alternative approach to the procedure just presented is described in \citet{Calanog14}, where {\sc galfit} and {\sc gravlens} \citep{Keeton01} are used iteratively to model the lens in the near-IR observed-frame. The surface mass density of the lens is assumed to be described by a singular isothermal ellipsoid \citep[SIE;][]{Kormann94}. The background source is assumed to comprise one or more components with S\'ersic light profiles \citep{Sersic68}. No foreground obscuration is considered. The best solution implies a complex background morphology (three components) and the following SIE parameters: 
$b=0.738^{+0.002}_{-0.001}$\arcsec\ (the Einstein radius), $\delta x = 0.027^{+0.002}_{-0.002}$\arcsec, $\delta y = 0.044^{+0.002}_{-0.003}$\arcsec, $\epsilon=0.208^{+0.005}_{-0.003}$ (the ellipticity), $\theta = -51.0^{+0.5}_{-0.4}\deg$. The fit quality, as assessed via $\chi^2$-statistics, is $\chi^2/\nu = 5452/2097 = 2.6$.


\subsubsection{Source-plane reconstruction}

\begin{figure*}
\centering
\includegraphics[scale=0.65]{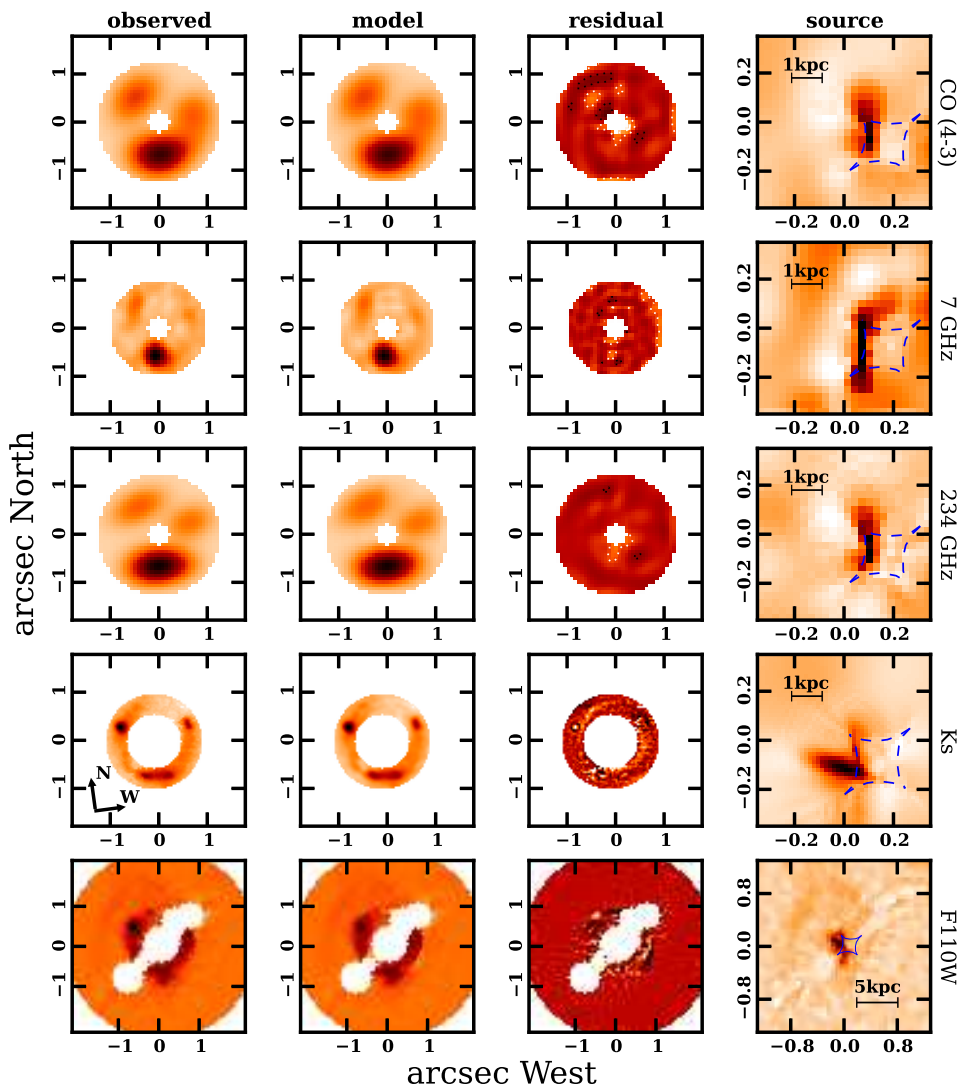}
\caption{The multi-wavelength source reconstruction of H1429$-$0028: CO\,(J:4$\to$3) (uniform weight, top row); 7-GHz continuum (Briggs weight, $\rm{robust}=5$, top middle row); 234-GHz continuum (uniform weight, middle row); $K_{\rm s}$-band (bottom middle row); and $F110W$-band (bottom row). Left column shows the observation data, while the second column shows the model image-plane. The third column shows the residuals with a scale range from $-4\sigma$ to $4\sigma$. Contours show $-3\sigma$ (solid white), $-1\sigma$ (dotted white), $1\sigma$ (dotted black), and $3\sigma$ (solid black) levels. The fourth column shows the source-plane reconstruction, with the caustic overlaid, and the physical scale given by the errorbar. The $K_{\rm s}$-band frames have a slight tilt shown by the two arrows in the left column.}
\label{fig:recon}
\end{figure*}

The best-fit parameter set was used to also reconstruct the source-planes of the 243\,GHz, $K_{\rm s}$-band, and $F110W$-band continua emissions. All the reconstructions are shown in Fig.~\ref{fig:recon}. While the 243-GHz continuum map is nicely modelled (no emission peaks above 1.6$\sigma$), the algorithm still struggles to fit knot C in the optical rest-frame imaging. This is also reported by \citet{Calanog14}, where even after considering a complex morphology, the fit was still poor. Here, the SLI method analysis is not limited to \emph{a priori} background morphology and different scenarios of foreground obscuration are considered (based on the {\sc galfit} foreground light profile, see Sec.~\ref{sec:fbsed}). Still, the fit is poor. Hence, one concludes that either the foreground obscuration is not properly accounted for (e.g., due to clumpiness), or the surface mass density models adopted in both works fail to explain the background morphology at high spatial resolution ($\lesssim0.2\arcsec$).

The background source morphology at long wavelengths is dominated by extended north-south (NS) emission along the fold and reaching the cusps. There is emission dispersion to the north of the north-west cusp and to the east of the fold. The latter coincides with the dominant emission in the $K_{\rm s}$-band with an approximately east-west (EW) direction. Notice that the NS feature is also observed in the $K_{\rm s}$-band, even though significantly fainter. The two features show a position angle of $\sim80\,$deg between each other and seem to be two distinct components. We take this as evidence for a merger system. Such a scenario may explain the north-eastern arc-like outer feature observed in the $F110W$ band with a length of a few tens of kpc in the source plane.

In \citet{Calanog14}, the background source is found to be composed of two small sources (effective semi-major axis of $\sim0.03\arcsec$) and a larger ($\sim0.18\arcsec$) third component with a north-south position falling to the east side of the caustic. In terms of surface brightness, the two smaller sources dominate and fall along the position of the EW feature referred to above. Hence, neither approach retrieves an acceptable fit to the near-IR dataset, we consider that the background morphology at those wavelengths is fairly consistent between the two independent results.

\begin{figure}
\centering
\includegraphics[scale=0.2]{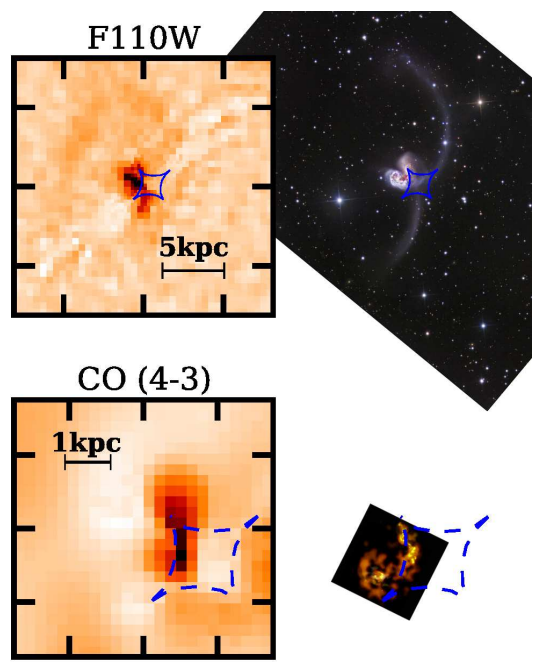}
\caption{Using the Antenn\ae\ galaxy merger as a toy model to help visualise the background galaxy of H1429-0028. The two upper images show the $F110W$ source-plane reconstruction from Fig.~\ref{fig:recon} (left hand-side) and the Antenn\ae\ galaxies as seen in the optical (right hand-side). The two bottom images show the CO\,(J:4$\to$3) velocity-integrated flux source-plane reconstruction from Fig.~\ref{fig:recon} (left hand-side) and the CO\,(J:J$^{\rm up}=1,2,3$) map in the Antenn\ae\ galaxies as observed by ALMA. \emph{Image credits: NASA, ESA and Ivo Saviane (upper-right picture), ALMA (ESO, NRAO, NJAO; bottom-right picture)}.}
\label{fig:antcomp}
\end{figure}

As for a toy model, we show in Fig.~\ref{fig:antcomp} the Antenn\ae\ galaxies for comparison. Although it is not a 100\% match, the resemblance is significant, explaining properties such as why the optical and mm frames are dominated by different components, and the presence of tidal tails appearing North-Eastward and southward to the caustic on, respectively, the $F110W$ imaging and the $F110W$ and 7-GHz imaging.

\subsubsection{Source dynamical analysis} \label{sec:dynamass}

\begin{figure*}
\includegraphics[scale=0.25,left]{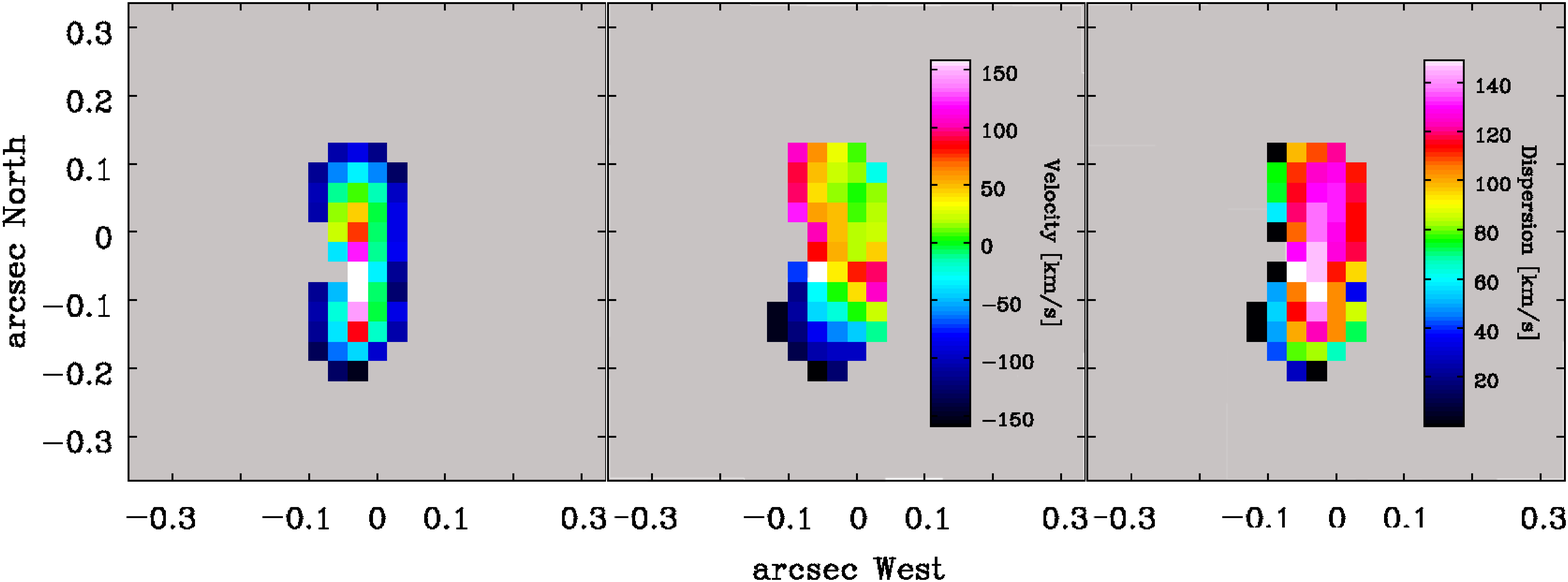}
\caption{The source-plane dynamical properties of H1429$-$0028. The left hand-side panel shows the velocity-integrated intensity map. The middle panel shows the velocity field, while the right hand-side panel shows the velocity dispersion map. Colour scales for the middle and right panels are given by the inset bar.}
\label{fig:dyn}
\end{figure*}

The dynamical analysis is, at this point, limited to the north-south component. Future optical integral field spectroscopy or deeper ALMA observations are required to study the east-west component. Applying the best-fit lens model to the CO\,(J:$4\to3$) cube allows one to study the source dynamics in the source plane. Fig.~\ref{fig:dyn} shows the CO\,(J:$4\to3$) moment maps: velocity-integrated line flux, velocity field, and velocity dispersion. It is clear that the southern emission is predominantly blueshifted, while that in the north is predominantly redshifted.

In order to estimate the dynamical mass of the background source, we consider the `isotropic virial estimator':
\[ \rm{M_{dyn} = 2.8\times10^5 ~ (\Delta v_{\rm FWHM})^2~r_{1/2}}, \]
where [$\rm{M_{dyn}}$] = M$_\odot$, $\Delta v_{\rm FWHM}$ is the CO\,(J:4$\to$3) FWHM (481$\pm$13\,km\,s$^{-1}$), and $r_{1/2}$ is the half-light radius.

With such disturbed source-plane morphology, we avoid fitting a light-profile (e.g.\ S\'ersic), and adopt an alternative method to estimate $r_{1/2}$. First, the source centre ($x_c,y_c$) is found by minimising the second-order moment of the source pixels:
\[ \rm{M_{total}} = \sum\limits_i^n~f_i~[(x_i-x_c)^2 + (y_i-y_c)^2], \]
where $f_i$, $x_i$, and $y_i$ are, respectively, the flux and coordinates of each pixel. The half-light radius was then considered to be equal to the maximum extension in respect to the estimated source centre among the pixels comprising half the source flux. The uncertainties were found via bootstrapping, i.e., the pixel flux values were shifted within $\pm$r.m.s., and $r_{1/2}$, $x_c$, and $y_c$ were recomputed for a total of 10,000 iterations. The estimated half-light radius is found to be $r_{1/2} = 0.90\pm0.26\,$kpc, and the dynamical mass to be $5.8\pm1.7\times10^{10}\,$M$_\odot$. The discussion continues in Sec.~\ref{sec:dynsedmass}, after stellar, dust and inter stellar medium (ISM) gas masses are estimated.

\subsubsection{Magnification factor} \label{sec:magn}

Table~\ref{tab:magn} shows the estimated magnifications depending on wavelength and source-plane region. The latter is addressed in the columns $\mu_{\rm TOT}$, $\mu_{50}$, and $\mu_{10}$ referring, respectively, to the ratio between the total image flux and the total source-plane flux, and the magnification of the brightest region in the source-plane that contains 50\% and 10\% of the total source-plane flux. Considering these, the spatial differential magnification is clear, with differences of up to a factor of $\sim$4. This is unsurprising given the spatial extent of the source and its proximity to the caustic. Hereafter, the adopted value for magnification will be $\mu_{\rm TOT}$.

\begin{table}[!ht]
\caption{The multi-wavelength magnification of H1429$-$0028.}
\label{tab:magn}
\centering
\begin{tabular}{crrr}
\hline\hline
Data & $\mu_{\rm TOT}$\tablefootmark{a} & $\mu_{50}$\tablefootmark{b} & $\mu_{10}$\tablefootmark{c} \\
\hline
\smallskip
\emph{HST} $F110W$ & $7.9\pm0.8$ & $11.0\pm0.9$ & $7.3\pm0.9$ \\
Keck $K_{\rm s}$ & $8.9\pm0.7$ & $11.0\pm0.7$ & $11.2\pm0.7$ \\
ALMA CO\,(J:4$\to$3) & $9.7\pm0.7$ & $13.9\pm0.9$ & $25.0\pm1.9$ \\
ALMA 234\,GHz     & $10.8\pm0.7$ & $14.3\pm0.8$ & $26.0\pm2.0$ \\
JVLA 7\,GHz &  $5.2\pm0.5$ & $11.6\pm1.1$ & $20.2\pm1.8$ \\
\hline
\end{tabular}
\tablefoot{
\tablefoottext{a}{The ratio between the total image flux and the total source-plane flux.}\\
\tablefoottext{b}{The magnification of the brightest region in the source-plane that contains 50\% of the total source-plane flux.}\\
\tablefoottext{c}{The magnification of the brightest region in the source-plane that contains 10\% of the total source-plane flux.}\\
}
\end{table}

\subsubsection{Stellar mass contribution to the lens effect}

In Section~\ref{sec:fbsed} we estimate the foreground stellar mass to be $2.8_{-1.2}^{+2.0}\times10^{10}\,\rm{M_\sun}$. Adopting the fraction of light in the $K_{\rm s}$ band within the average Einstein radius (57\%) to be a proxy of the fraction of the stellar mass within that same radius ($1.60_{-0.69}^{+1.1}\times10^{10}\,\rm{M_\sun}$), one can estimate the stellar mass contribution to the lens effect. From the lens analysis and integrating Equation~\ref{eq:kappa} over theta and radius, we know that the total mass within the average Einstein radius is $\rm{M(<\theta_E)}=8.13_{-0.41}^{+0.33}\times10^{10}~$M$_\odot$. Hence, the stellar mass contribution to the deflection effect is $19.7_{-8.5}^{+14}\%$.

\subsection{Foreground and background SEDs} \label{sec:fbsed}

As expected for a gravitational lens system such as H1429$-$0028, the SED is actually a combination of two individual SEDs, and their deblending is required to study each galaxy separately. The spatially resolved photometry of the two galaxies in the $F110W$, $H$, and $K_{\rm s}$ bands, and ALMA observations indicate that, down to the data sensitivity, the background system is the sole contributor at least in the mm spectral range. Given the lack of spatially resolved photometry, we avoid working with best-fit solutions and consider instead flux probability distribution functions (PDFs). To obtain these, we utilise the {\sc magphys}\footnote{www.iap.fr/magphys} software \citep{daCunha08}. This code considers the latest version of the \citet{Bruzual03} stellar population synthesis code, where the new prescription by \citet{Marigo07} for the thermally pulsating asymptotic giant branch evolution of low- and intermediate-mass stars is considered.

We first remove foreground light from the total SED. The foreground rest-frame optical spectral range is traced by the high-resolution imaging in the $F110W$, $H$, and $K_{\rm s}$ bands. To help constrain the obscuration at short-wavelengths, we consider a $g$-band upper-limit 3$\sigma$ away from the total-flux detection. Also, the non-detection at mm wavelengths provides an upper-limit (at the 3$\sigma$ level) with which to better constrain that region of the foreground SED. In order to take into account errors in the multi-waveband analysis (i.e., mismatched aperture sizes and absolute zero points, and calibration errors), we add in quadrature 0.1\,magnitudes ($\sim$9\% of the flux) to the photometric error. This procedure yields a flux PDF for each band\footnote{The use of upper-limits and the extraction of flux PDF was possible after changing the standard {\sc magphys} code publicly available.}. These PDFs were used to determine the amount of foreground flux to remove from total photometry flux. The difference between the two in each band gives the background flux.

Before one proceeds to analyse the resulting background SED, one has to correct for the possible foreground extinction affecting knots A, B, and D at short wavelengths (Fig.~\ref{fig:optnirCO}). We thus consider three scenarios: (i) there is no foreground extinction, (ii) the extinction is linearly proportional to or (iii) weighted on pixel flux of the foreground light-profile model obtained with {\sc galfit}. The difference between scenarios (ii) and (iii) is that extinction will be more centrally concentrated in (iii). Scenarios (ii) and (iii) can be translated into the following equations:
\begin{equation}
  \rm{(ii)} ~ e^{-\tau_\lambda}_{\rm i} = f_{\rm i}\frac{e^{-\tau_\lambda}}{\overline{f}}
\end{equation}
\begin{equation}
  \rm{(iii)} ~ e^{-\tau_\lambda}_{\rm i} = f_{\rm i}\frac{e^{-\tau_\lambda}}{\frac{\sum{f_{\rm i}^2}}{\sum{f_{\rm i}}}} \times \frac{f_{\rm i}}{\frac{\sum{f_{\rm i}^2}}{\sum{f_{\rm i}}}} = e^{-\tau_\lambda} \left( \frac{f_{\rm i}}{\frac{\sum{f_{\rm i}^2}}{\sum{f_{\rm i}}}} \right)^2
\end{equation}

\noindent where the \emph{i} index refers to a given pixel at $>3\sigma$, \emph{$f_i$} is the pixel flux, $\overline{f}$ and $\frac{\sum{f_{\rm i}^2}}{\sum{f_{\rm i}}}$ are, respectively, the foreground light-model flux average and weighted average of pixels at $>3\sigma$, and $\tau_\lambda$ is the absorption optical depth at a given waveband. The latter is obtained with equations (3) and (4) from \citet{daCunha08}, where $\tau_V$ is given in this work by the {\sc magphys} analysis of the foreground SED, which yields $\tau_V=3.39^{+0.86}_{-0.96}$. The assumed background morphology is that observed in the Keck AO $K_{\rm s}$-band. The final adopted flux estimate is the average between the maximum and minimum values (including error) from the three scenarios, with an error equal to the maximum deviation from the average.

The `continuum' data point from MAMBO2 was left out from the analysis due to line contamination. Considering the MAMBO2 response curve\footnote{http://www.astro.uni-bonn.de/$\sim$bertoldi/projects/mambo/manuals.html}, the ALMA estimates for the continuum level, line-flux estimates for CO($4\to3$) and C\,I($\rm{^3P_1}\to^3\rm{P_0}$), and Z-SPEC for CO($5\to4$), and a spectral slope $\alpha=3.89\pm0.09$ (Sec.~\ref{sec:almacont}), one should observe a MAMBO2 flux of $\sim10.0\pm0.14\,$mJy, which is more in agreement with the actual observed value ($10.3\pm2.5\,$mJy). The ALMA band-3 continuum estimate (rest-frame 216\,GHz) was also discarded given the evidence for free-free emission contamination \citep[see also][]{Condon92,Thomson12,Clemens13} during the analysis.

We also attempt to correct for differential magnification by demagnifying the bands shortward the $K_{\rm s}$-band (inclusive) by a factor of $8.9\pm0.7$, by $9.85\pm1.65$ the bands shortward of 1.28\,mm (exclusive), and by $10.8\pm0.7$ the 234-GHz continuum flux (Table~\ref{tab:magn}).

Finally, given the nature of the source, the standard models accompanying {\sc magphys} have shown limitations to cover the necessary physical parameter space, as described by \citet{Rowlands14}; hence, for the background source, we have adopted the models presented in that same work, which are better suited for more extreme star-forming systems.

\begin{figure}
\centering
\includegraphics[scale=0.45]{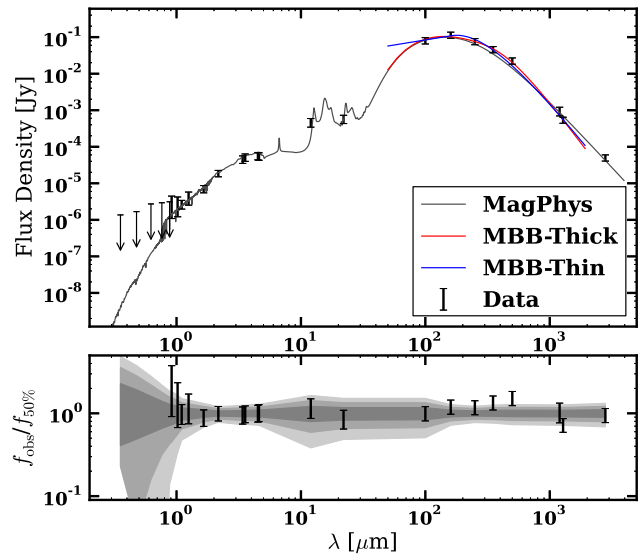}
\caption{Comparing the input flux data points (errorbars) and the predicted fluxes by {\sc magphys} (grey regions referring to 1$\sigma$, 2$\sigma$, and 3$\sigma$ confidence intervals in the bottom panel). The best fit model from the {\sc magphys} analysis is shown as a grey solid line. The best-fit models from the modified black body fitting are shown as red (optically-thick case) and blue (optically-thin) solid lines. The bottom panel shows the flux ratios with respect to the 50\% quartile of the flux PDF at a given band.}
\label{fig:sedfit}
\end{figure}

\begin{figure*}
\centering
\includegraphics[scale=0.45]{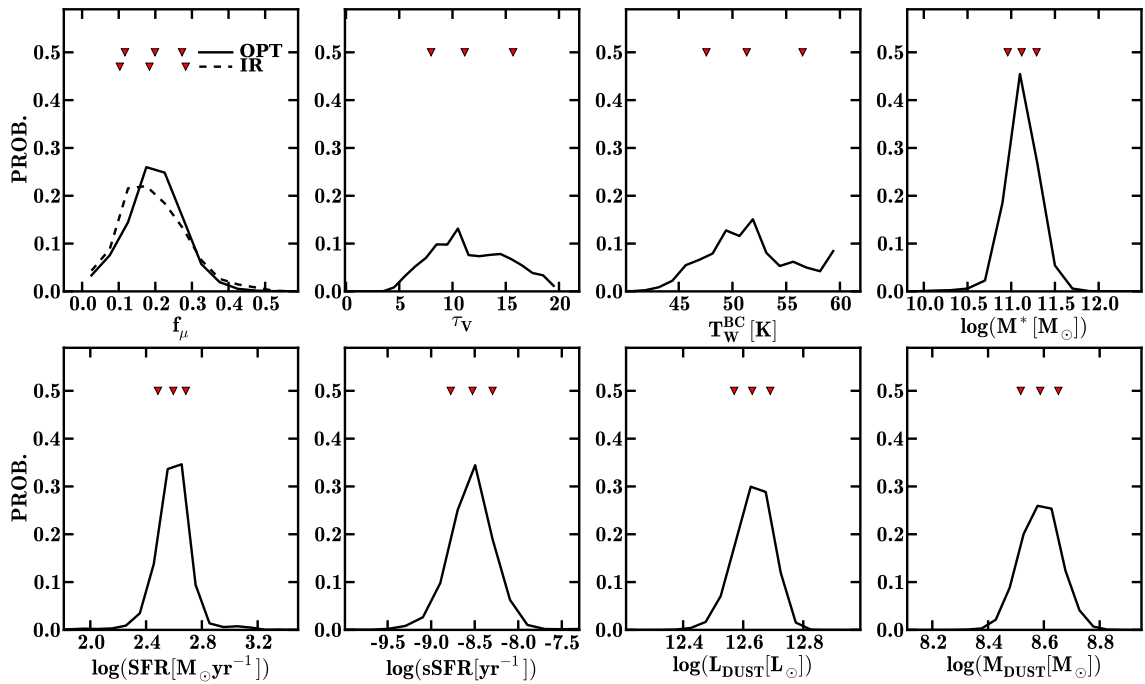}
\caption{The background SED fit analysis results. Each panel shows the PDF of a given physical parameter (top left: energy fraction absorbed by the ISM as estimated from stellar-dominated, OPT, or dust-dominated photometry, IR; top middle left: total effective $V$-band optical depth seen by stars in birth clouds; top middle right: warm dust temperature in birth clouds; top right: stellar mass; bottom left: star-formation rate; bottom middle-left: specific star-formation rate; bottom middle right: dust luminosity; bottom right: dust mass). The red inverted triangles indicate the 16$^{\rm th}$, 50$^{\rm th}$, and 84$^{\rm th}$ percentiles.}
\label{fig:pdfs}
\end{figure*}

Considering the above assumptions, Table~\ref{tab:mwlph} shows the observed and model predicted fluxes for both foreground and background systems. For the latter, the predicted fluxes are compared to input flux values (after foreground removal and flux demagnification) in Fig.~\ref{fig:sedfit}. Table~\ref{tab:magphys} shows the SED fit results of the physical properties for both fore and background SEDs. The PDFs of the background physical parameters are shown in Fig.~\ref{fig:pdfs}. The ISM dust temperature is poorly constrained, hence not shown.

\begin{table*}
\caption{Multi-wavelength photometry of H1429$-$0028.}
\label{tab:mwlph}
\centering
\begin{tabular}{ccrrr}
\hline\hline
Survey/ & Filter & Total & Foreg. & Backg.\tablefootmark{a} \\
Facility &  & [$\mu$Jy] & [$\mu$Jy] & [$\mu$Jy] \\
\hline
\smallskip
SDSS & $u$ & $<$4.9 & $0.80_{-0.45}^{+0.95}$ & $0.004_{-0.003}^{+0.006}$ \\
\smallskip
 & $g$ & 5.9$\pm$0.8 & $4.3_{-1.7}^{+2.7}$ & $0.024_{-0.012}^{+0.022}$ \\
\smallskip
 & $r$ & 20.5$\pm$1.1 & $15.1_{-4.0}^{+5.1}$ & $0.11_{-0.05}^{+0.07}$ \\
\smallskip
 & $i$ & 35.3$\pm$1.9 & $26.8_{-5.0}^{+5.7}$ & $0.38_{-0.13}^{+0.15}$ \\
\smallskip
 & $z$ & 61.9$\pm$7.4 & $42.4_{-5.6}^{+5.9}$ & $1.19_{-0.33}^{+0.33}$ \\
\smallskip
VIKING & $Z$ & 52.7$\pm$3.1 & $40.7_{-5.6}^{+5.9}$ & $1.11_{-0.32}^{+0.32}$ \\
\smallskip
 & $Y$ & 78.8$\pm$5.9 & $59.0_{-6.1}^{+6.1}$ & $1.80_{-0.40}^{+0.40}$ \\
\smallskip
 & $J$ & 133.9$\pm$6.7 & $102.1_{-4.8}^{+4.1}$ & $3.4_{-0.5}^{+0.5}$ \\
\smallskip
 & $H$ & 192$\pm$12 & 134.8$\pm$8.4 & 6.4$\pm$0.6\tablefootmark{b} \\
\smallskip
 & $K_{\rm s}$ & 380$\pm$12 & 224.7$\pm$6.9 & 17.4$\pm$1.5\tablefootmark{b} \\
\smallskip
\emph{HST} & $F110W$ & 98.8$\pm$4.2 & 78.1$\pm$3.5 & 2.6$\pm$0.3\tablefootmark{b} \\
\hline\hline
Survey/ & Filter & Total & Foreg. & Backg. \\
Facility &  & [mJy] & [mJy] & [mJy] \\
\hline
\emph{Spitzer} & $3.6\,\mu$m & 0.614$\pm$0.003 & $0.130_{-0.023}^{+0.031}$ & $0.052_{-0.005}^{+0.004}$ \\
\smallskip
 & $4.5\,\mu$m & 0.673$\pm$0.004 & $0.136_{-0.029}^{+0.042}$ & $0.056_{-0.005}^{+0.005}$ \\
\smallskip
\emph{WISE} & $3.4\,\mu$m & 0.558$\pm$0.014 & $0.131_{-0.022}^{+0.029}$ & $0.047_{-0.004}^{+0.004}$ \\
\smallskip
 & $4.6\,\mu$m & 0.653$\pm$0.020 & $0.130_{-0.028}^{+0.038}$ & $0.053_{-0.005}^{+0.005}$ \\
\smallskip
 & $12\,\mu$m & 5.39$\pm$0.15 & $0.78_{-0.48}^{+0.84}$ & $0.40_{-0.08}^{+0.08}$ \\
\smallskip
 & $22\,\mu$m & 6.22$\pm$0.76 & $0.51_{-0.31}^{+0.59}$ & $0.67_{-0.11}^{+0.12}$ \\
\smallskip
\emph{Herschel} & $100\,\mu$m & 821$\pm$28 & $23_{-12}^{+20}$ & $80_{-13}^{+14}$ \\
\smallskip
 & $160\,\mu$m & 1164$\pm$32 & $33_{-15}^{+23}$ & $94.8_{-8.9}^{+8.7}$ \\
\smallskip
 & $250\,\mu$m & 778$\pm$6 & $23_{-10}^{+15}$ & $64.4_{-4.9}^{+4.6}$ \\
\smallskip
 & $350\,\mu$m & 467.0$\pm$7.0 & $11.4_{-5.3}^{+7.7}$ & $34.0_{-2.7}^{+2.9}$ \\
\smallskip
 & $500\,\mu$m & 227.0$\pm$8.0 & $4.6_{-2.2}^{+3.2}$ & $14.7_{-1.3}^{+1.3}$ \\
\smallskip
IRAM-30 & 1.2\,mm & 10.3$\pm$2.5 & $0.23_{-0.11}^{+0.18}$ & $0.910_{-0.086}^{+0.081}$ \\
\smallskip
ALMA & 1.28\,mm & 5.86$\pm$0.99 & $0.18_{-0.09}^{+0.15}$ & 0.543$\pm$0.098\tablefootmark{c} \\
\smallskip
 & 2.8\,mm & 0.54$\pm$0.11 & $0.0014_{-0.0007}^{+0.0011}$ & $0.0525_{-0.0055}^{+0.0056}$ \\
\smallskip
JVLA & 7\,GHz & 0.91$\pm$0.08 & $<$0.03 & 0.175$\pm$0.023\tablefootmark{c} \\
\hline
\end{tabular}
\tablefoot{Upper limits are set at the $3\sigma$ level. \\
\tablefoottext{a}{Intrinsic (demagnified) fluxes.} \\
\tablefoottext{b}{Estimated via direct analysis of high-resolution \emph{HST F110W}, Keck AO $H$ and $K_{\rm s}$ imaging and demagnified by $\mu=7.9\pm0.8$ in the former and $\mu=8.9\pm0.7$ in the Keck bands.} \\
\tablefoottext{c}{Assumed to be equal to the total flux and demagnified by $\mu=10.8\pm2$ at 1.28\,mm and $\mu=5.2\pm0.5$ at 7\,GHz (Table~\ref{tab:magn}).}
}
\end{table*}

\begin{table*}
\caption{{\sc magphys} SED analysis.}
\label{tab:magphys}
\centering
\begin{tabular}{crrrrrrrrr}
\hline\hline
Source & $f_\mu$\,(SFH/IR)\tablefootmark{a} & $\tau_V$ & T$^{ISM}$ & T$^{BC}$ & M* & SFR & sSFR & L$_{\rm dust}$ & M$_{\rm dust}$ \\
 & & & [K] & [K] & [10$^{10}$\,M$_\sun$] & [M$_\sun$\,yr$^{-1}$] & [10$^{-10}$\,yr$^{-1}$] & [10$^{11}$\,L$_\sun$] & [10$^{8}$\,M$_\sun$] \\
\hline
\smallskip
Foreg. & 0.81$_{-0.18}^{+0.14}$& 3.39$_{-0.96}^{+0.86}$ & 22.1$_{-3.0}^{+2.1}$ & 45.5$_{-10}^{+9.8}$ & 2.8$_{-1.2}^{+2.0}$ & 1.2$_{-0.9}^{+2.5}$ & 0.43$_{-0.36}^{+1.3}$ & 0.44$_{-0.21}^{+0.34}$ & 0.44$_{-0.23}^{+0.45}$ \\
\smallskip
& 0.79$_{-0.19}^{+0.13}$ & \\
\smallskip
Backg. & 0.199$_{-0.082}^{+0.074}$ & 11.2$_{-3.2}^{+4.5}$ & \tablefootmark{b}28.88$_{-2.1}^{+0.90}$ & 52.4$_{-2.8}^{+3.6}$ & 13.2$_{-4.1}^{+6.3}$ & 394$_{-88}^{+91}$ & 30$_{-13}^{+21}$ & 42.7$_{-5.5}^{+6.3}$ & 3.86$_{-0.58}^{+0.62}$ \\
\smallskip
& 0.184$_{-0.081}^{+0.099}$ & \\
\hline
\end{tabular}
\tablefoot{The main value refers to the 50\% percentile, while the errors refer to the deviation to the 16$^{\rm th}$ and 84$^{\rm th}$ percentiles.\\
\tablefoottext{a}{The values in the first and second rows refer to the energy fraction absorbed by the ISM as estimated from stellar-dominated ($f_\mu^{\rm SFH}$) and dust-dominated ($f_\mu^{\rm IR}$) photometry.}\\
\tablefoottext{b}{The T$^{ISM}$ PDF for the background source does not reach a peak, so value should not be considered reliable.}
}
\end{table*}

Although the fit is generally good, there is a slight tension in the 350-$\mu$m, 500-$\mu$m and 1.28-mm bands. The deviation in the \emph{Herschel} bands may be, respectively, assigned to emission from [C\,{\sc ii}] and CO~(J$^{\rm up}=9-12$). For instance, considering the [C\,{\sc ii}]-to-FIR relation from \citet{DiazSantos13}, we estimate a [C\,{\sc ii}] flux contribution to the 350-$\mu$m band of $8.2_{-4.7}^{+11.3}\%$ \citep[see also][]{Smail11}. Such an effect, however, does not explain the overestimate at 1.28\,mm, even though just at a $\sim2\sigma$ level.


Hence, we have also considered the algorithm which fits modified black-body models to photometry data \citep[{\sc emcee}\footnote{https://github.com/aconley/mbb\_emcee}, see, for instance,][]{Riechers13, Fu13} using an affine invariant Markov chain Monte Carlo method \citep{Foreman12}. The difference to {\sc magphys} is that, although {\sc emcee} considers only one emission component, the parameter range is not limited to input models and the optically-thick scenario can also be explored. Table~\ref{tab:greyb} summarises the results and Fig.~\ref{fig:sedfit} shows the best-fit models. In order to compute these values, it was necessary to limit $\beta$ below 3, the temperature to observed-frame 100\,K, and, in the optically-thick case, $\lambda_0$ below observed-frame 2000\,$\mu$m. While it is not straightforward to compare the success between the two codes based of the $\chi^2$ value (due to the constrain in certain parameters), both scenarios improve the fit to the FIR-to-mm spectral range. Interestingly enough, the dust mass and IR luminosity values are consistent within the errors with those obtained with {\sc magphys}.

\begin{table}[!ht]
\caption{The FIR-to-mm properties of H1429$-$0028.}
\label{tab:greyb}
\centering
\begin{tabular}{crrrrrrrr}
\hline\hline
Case & thin & thick \\
\smallskip
$T$ [K] & $35.9_{-4.3}^{+4.3}$ & $73_{-23}^{+13}$ \\
\smallskip
$\alpha$\tablefootmark{a} & $7.6_{-6.5}^{+7.5}$ & $8.3_{-5.6}^{+6.6}$ \\
\smallskip
$\beta$\tablefootmark{b} & $2.14_{-0.24}^{+0.23}$ & $2.25_{-0.65}^{+0.61}$ \\
\smallskip
$\lambda_0$ [$\mu$m]\tablefootmark{c} & --- & $779_{-405}^{+986}$ \\
M$_{\rm dust}$ $\rm{\left[10^8\,M_\sun\right]}$ & $5.2_{-2.6}^{+1.9}$ & $4.9_{-3.5}^{+1.8}$ \\
\smallskip
L$_{\rm 8-1000~\mu m}$ $\rm{\left[10^{12}\,L_\sun\right]}$ & $6.2_{-2.2}^{+2.5}$ & $4.9_{-3.1}^{+1.7}$ \\
\smallskip
$\nu$\tablefootmark{d} & 2 & 1 \\
\smallskip
$\chi^2$ & 1.94 & 0.62 \\
\hline
\end{tabular}
\tablefoot{Photometry analysis with {\sc emcee} using \emph{Herschel} 100--500-$\mu$m and ALMA 1.28-mm data only. Errors are $\pm1\sigma$. The {\sc emcee} analysis considers a covariance matrix to account for flux calibration issues and uncertainty.\\
\tablefoottext{a}{The mid-IR power-law index.}\\
\tablefoottext{b}{The extinction curve power-law index. $\beta$ was limited to values below 3.}\\
\tablefoottext{c}{Wavelength at which optical depth equals unity. $\lambda_0$ was limited to values below 2000\,$\mu$m.}\\
\tablefoottext{d}{Number of degrees of freedom. This is the number of photometric data points used (six), minus the number of parameters to fit in each case, four and five, respectively, for the optically thin and thick cases ($T$, $\alpha$, $\beta$, normalisation, $\lambda_0$).}
}
\end{table}

\subsection{Radio-FIR correlation and SFRs}

The direct comparison between of mm and cm imaging allows us to infer the radiation mechanisms responsible for both emissions. Specifically the ratio between the 8---1000\,$\mu$m and 1.4-GHz fluxes, the $q_{\rm TIR}$ parameter, has been frequently used to distinguish star-forming from AGN dominated regions, with a value of $2.64\pm0.26$ being characteristic of local star-forming galaxies with no signs of AGN activity \citep{Bell03}\footnote{\citet{Yun01} define the same parameter with reference to the 42.5--122.5\,$\mu$m spectral range. By doing so, normal star-forming galaxies are expected to have $q = 2.34\pm0.26$ \citep{Bell03}.}. In \citet{Bell03}, this parameter is defined as
\[ q_{\rm TIR}=\log_{10}\left(\frac{\rm{TIR}}{3.75\times10^{12}\,\rm{W\,m^{-2}}}\right) - \log_{10}\left(\frac{S_{1.4\,\rm{GHz}}}{\rm{W\,m^{-2}\,Hz^{-1}}}\right), \]
\noindent where TIR is the total 8--1000\,$\mu$m IR flux in $\rm{W\,m^{-2}}$, and $S_{1.4\,\rm{GHz}}$ is the 1.4-GHz flux density in $\rm{W\,m^{-2}\,Hz^{-1}}$. We convert the observed 7-GHz flux densities to rest-frame 1.4-GHz flux densities assuming a power-law index of 0.8$\pm$0.2 ($f\propto\nu^{-0.8}$) characteristic of synchrotron radiation \citep[e.g.][]{Ibar10rad, Thomson14}. We consider the different magnifications of $10.8\pm0.7$ and $5.2\pm0.5$, respectively, for the IR and radio spectral regimes. In H1429$-$0028, we find $q_{\rm TIR}=1.9_{-1.2}^{+1.1}$, which is consistent with the value found for normal galaxies within $1\sigma$.

Finally, we estimate IR and radio SFRs by assuming the IR and radio luminosity-to-SFR calibrations proposed by \citet[][see also \citealt{Kennicutt98}]{Bell03}, which account for the contribution from old stellar populations:
\begin{equation}
  \rm{SFR[M_\odot~yr^{-1}]} \equiv 
  \begin{cases}
    1.57\times10^{-10}~L_{\rm TIR}~\left(1+\sqrt{10^9/L_{\rm TIR}}\right) \\
    5.52\times10^{-22}~L_{\rm 1.4\,GHz}
  \end{cases}
\end{equation}
where $[L_{\rm 1.4\,GHz}] = \rm{W~Hz^{-1}}$, $[L_{\rm TIR}]=[L_{\rm 8-1000~\mu m}] = \rm L_\odot$. The intrinsic IR and radio SFRs are estimated to be, respectively, $3.9_{-2.0}^{+8.1}\times10^2$ and $9.3_{-6.5}^{+20}\times10^2$\,M$_\sun$\,yr$^{-1}$, where the error takes into account a factor of 2 due to the expected scatter in the relations \citep{Bell03}. Within 1$\sigma$, these estimates are in agreement with that obtained from the {\sc magphys} analysis.

\subsection{Molecular gas and ISM gas masses of the background galaxy} \label{sec:molecmass}

As previously mentioned, the available number of CO transitions is not enough to properly constrain the CO ladder, nor do we have a reliable CO\,(J:5$\to$4) flux measurement. Hence, in order to compute total molecular gas masses, one has to rely on certain assumptions or empirical statistical relations available in the literature.

It is clear from the detection of CS\,(J:10$\to$9) that the observed background $^{12}$CO emission is optically-thick. Nevertheless, assuming local thermodynamical equilibrium (LTE) and an optical thin transition, one can estimate a mass lower limit. We assume the nomenclature $X^{\rm thin}_{\rm CO}=\frac{\rm M_{H_2}}{\rm L'_{CO\,(J:1\to0)}}$, where

\[ X^{\rm thin}_{\rm CO} \sim 0.08 \left[ 
\frac{g_1}{Z} e^{-T_{\rm o}/T_{\rm k}}
\left( \frac{J(T_{\rm k})-J(T_{\rm bg})}{J(T_{\rm k})} \right)
\right] ^{-1} \]
\[ \times
\left(
\frac{[\rm{CO/H2}]}{10^{-4}}
\right) ^{-1}
{\rm \frac{M_\sun}{K~km~s^{-1}~pc^2}},
\]

\noindent with $T_{\rm o}=E_u/k_B\sim5.5$\,K, $J(T)=T_{\rm o}(e^{T_{\rm o}/T}-1)^{-1}$, $T_{\rm bg}=(1+z)T_{\rm CMB}=5.524$\,K (the temperature of the cosmic microwave background at $z=1.027$), $g_1=3$ (the degeneracy of level $n=1$), $Z\sim2T_{\rm k}/T_{\rm o}$ (the partition function), $T_{\rm k}$ is assumed to be equal to that estimated by {\sc emcee} in the optically-thin case ($35.9_{-4.3}^{+4.3}\,$K, Table~\ref{tab:greyb}), $[\rm{CO/H2}]=10^{-4}$ \citep[the CO abundance in typical molecular clouds or in a solar-metallicity environment,][]{BryantScoville96}, and Helium mass is already considered. These assumptions result in $X^{\rm thin}_{\rm CO} \sim 0.449_{-0.039}^{+0.040}$. Due to the absence of a CO(1$\to$0) observation, we base our CO\,(J:1$\to$0) line luminosity in that observed for CO\,(J:2$\to$1) assuming a conversion factor of ${\rm L'_{CO\,2\to1}}/0.92$ \citep{CarilliWalter13}. This yields a molecular mass lower limit of ${\rm M_{H_2}}>1.03_{-0.11}^{+0.11}\times10^{11}\,\mu^{-1}\,{\rm M_\sun}$. Following \citet{Ivison11}, an upper limit may be estimated assuming $X_{\rm CO}=5$ \citep[observed in giant molecular clouds,][]{Solomon87,SolomonBarrett91}, which yields ${\rm M_{H_2}}<1.15\pm0.08\times10^{12}\,\mu^{-1}\,{\rm M_\sun}$. Hence, based on the CO\,(J:2$\to$1) observations, we expect the intrinsic molecular mass to be in the range $1.80_{-0.29} < \rm{M_{H_2}/10^{10}\,[M_\sun]} < 4.08^{+0.66}$.

Recently, \citet{Narayanan12} proposed a simple relation between $X_{\rm CO}$\footnote{The nomenclature for $X_{\rm CO}$ we adopt here is that adopted in \citet{Narayanan12} for $\alpha_{\rm CO}$.} and CO and metallicity measurements:
\[ X_{\rm CO} = \frac{10.7\times \langle W_{\rm CO}\rangle^{-0.32}}{Z'^{0.65}}, \]
where $\langle W_{\rm CO}\rangle$ is the luminosity-weighted CO intensity, measured in K\,km\,s$^{-1}$, and $Z'$ is the metallicity divided by the Solar metallicity. Assuming $Z'=0.50_{-0.25}^{+0.50}$, the relation yields $X_{\rm CO} = 6.6_{-4.3}^{+2.2}$. The errors are still consistent with the range we adopted previously. However, either assumption implies a large range of $M_{\rm H_2}$.

An alternative to using CO emission to estimate $M_{\rm H_2}$ is to use the forbidden fine-structure transitions of neutral carbon ([C\,{\sc i}]). The critical density of both [${\rm C\,I}$] and $^{12}$CO are $n\sim10^3\,{\rm cm^{-3}}$.  Also, [C\,{\sc i}] traces only molecular gas, as a result of being insensitive to the presence of atomic or ionised gas.  However, here too, one must assume an optically thin [C\,{\sc i}] line in LTE in order to estimate [C\,{\sc i}] masses as
\[ M_{\rm C\,I}=5.706\times 10^{-4} Q(T_{\rm ex})\frac{1}{3}e^{23.6/T_{\rm ex}} L'_{\rm C\,I(^3P_1\to^3P_0)}\]
\noindent where $[M_{\rm C\,I}]=M_\sun$, $[L'_{\rm C\,I(^3P_1\to^3P_0)}]=\rm{K~km~s^{-1}~pc^2}$, and $Q(T_{\rm ex})=1+3e^{-T_1/T_{\rm ex}}+5e^{-T_2/T_{\rm ex}}$ is the [C\,{\sc i}] partition function, with $T_1=23.6\,$K and $T_2=62.5\,$K being the energies above the ground state for the [C\,{\sc i}($^3P_1\to^3P_0$)] and [C\,{\sc i}($^3P_2\to^3P_1$)] lines, respectively. The [C\,{\sc i}($^3P_2\to^3P_1$)] line, however, is redshifted to ALMA band 8, which was not available during Cycle~0, preventing an estimate of $T_{\rm ex}$. Hence, we again assume the value estimated by {\sc emcee} in the optically-thin case ($35.9_{-4.3}^{+4.3}\,$K, Table~\ref{tab:greyb}). For $L'_{\rm C\,I(^3P_1\to^3P_0)} = 4.15\pm 0.60\times 10^{10}\,\mu^{-1}\,\rm{K~km~s^{-1}~pc^2}$, we estimate a lower limit of $M_{\rm C\,I}>5.23_{-0.76}^{+0.76}\times10^7\,\mu^{-1}\,{\rm M_\sun}$. \citet{Weiss05} found, for a sample of three $z\sim2.5$ sources, a carbon abundance of $X[{\rm C\,I}]/X[{\rm H_2}]={\rm M_{C\,I}/(6 M_{H_2})}\sim5\times10^{-5}$, which is roughly double that found in our Galaxy \citep[$2.2\times10^{-5}$,][]{Frerking89}. We note that the reported $M_{\rm H_2}$ masses in \citet{Weiss05} were estimated based on CO emission assuming $X^{\rm thin}_{\rm CO}=0.8$, and the Carbon abundance in our Galaxy is likely not representative of that in H1429$-$0028. We assume the Galactic value provides an upper limit instead. Hence, assuming the range of $X[{\rm C\,I}]/X[{\rm H_2}]$ values, we expect the intrinsic molecular mass to be in the range $1.80_{-0.29} < M_{\rm H_2}/10^{10}\,[{\rm M_\sun}] < 4.08^{+0.66}$.

Finally, \citet{Scoville14} propose an empirical approach to estimate total ISM gas mass ($M_{\rm H\,I}+M_{\rm H_2}$) based on submm continuum emission. The relation is calibrated with a sample of local galaxies for which global $M_{\rm H\,I}$ and $M_{\rm H_2}$ estimates as well as submm observations exist. The reference wavelength is set at rest-frame 850\,$\mu$m, which traces the Rayleigh-Jeans tail of an SED. The relation is the following:
\[ M_{\rm ISM}=1.2\times10^4~D^2_{\rm L} \left(\frac{350}{\nu_{\rm obs}} \right)^{\beta} (1+z)^{-(1+\beta)}~S_{\nu_{\rm obs}}~\mu^{-1},\]
\noindent where $[M_{\rm ISM}]={\rm M_\sun}$, $[D_{\rm L}]={\rm Mpc}$, $[S_{\nu_{\rm obs}}]={\rm mJy}$, $[\nu_{\rm obs}]={\rm GHz}$ (350 corresponds to the frequency in GHz at 850\,$\mu$m), $\beta$ is the FIR--mm power-law index ($\beta=3.89\pm0.41$ in our case, Section~\ref{sec:almacont}). Hence, adopting our 1.28-mm flux density estimate of $5.86\pm0.99\,$mJy and a magnification of $\mu=10.8\pm0.7$, the estimated intrinsic ISM gas mass is $M_{\rm ISM} = 4.6\pm1.7\times 10^{10}\,{\rm M_\sun}$ \citep[with a conservative 25\% uncertainty added in quadrature due to the expected scatter of the adopted relation,][]{Scoville14}. This means a molecular-to-total gas mass ratio of $0.39_{-0.16} < M_{\rm H2}/M_{\rm ISM} < 0.89^{+0.11}$, a gas-to-baryonic mass fraction of $0.26_{-0.13}^{+0.15}$, and a depletion time of $\tau_{\rm SF}=M_{\rm ISM}/{\rm SFR}=117_{-51}^{+51}\,$Myr. This timescale is in agreement with that expected for the SMG phase \citep[$\sim$100\,Myr,][]{Greve04,Tacconi06,Tacconi08,Ivison11}. However, despite the evidence for a high-density and dusty environment (the detection of CS\,10$\to$9 and $M_{\rm dust}=3.86_{-0.58}^{+0.62}\times10^8\,{\rm M_\sun}$), which could make $H_2$ formation easier \citep[][and references therein]{Krumholz14}, such a short timescale and the fact that star formation is more directly related to molecular gas and not so much to total gas, or more specifically, neutral gas \citep[][and references therein]{Elmegreen11}, may imply a much shorter starburst phase, in the range $46_{-13}<M_{\rm H_2}/{\rm SFR}<104^{+29}$\,Myr, and a longer depletion time.

\subsection{Comparing dynamical and SED masses} \label{sec:dynsedmass}

In Sections~\ref{sec:dynamass}, \ref{sec:fbsed} and \ref{sec:molecmass}, we show that the expected background dynamical and baryonic masses are, respectively, $5.8\pm1.7\times10^{10}\,$M$_\odot$ and $17.8_{-4.4}^{+6.5}\times10^{10}\,$M$_\odot$. If, for the latter, one considers a $20\pm10$\% contribution from dark matter \citep{Gerhard01, Kassin06, Daddi10a}, $22.3_{-5.8}^{+8.3}\times10^{10}\,$M$_\odot$, there is a significant tension between the mass estimates obtained via the dynamical and SED data. The point driving this discrepancy is the fact that the dynamical information traced by the CO\,(J:4$\to$3) emission is dominated by the north-south component, whereas the SED information comes from the system as a whole, thus including the east-west component. As a result, one may estimate the dynamical mass of the east-west component by assuming it is the difference between the SED-derived total mass and the north-south component dynamical mass. Such an assumption implies a dynamical mass of $16.5_{-6.0}^{+8.5}\times10^{10}\,$M$_\odot$ for the east-west component. This means we may be witnessing a 1:2.8$_{-1.5}^{+1.8}$ intermediate-to-major merger at $z=1.027$.

\section{Conclusions} \label{sec:conclusions}

This work focus on a \emph{Herschel} 500-$\mu$m-selected source, HATLAS\,J142935.3$-$002836 (H1429$-$0028), a candidate lensed galaxy. The lensing scenario is confirmed with the help of multi-wavelength, high-resolution imaging (Fig.~\ref{fig:optnirCO}) which reveals a foreground edge-on disk galaxy surrounded by an almost complete Einstein ring.

Optical and FIR spectroscopy allow to measure, respectively, a foreground redshift of $z_{\rm sp}=0.218$ and a background redshift of $z_{\rm sp}=1.027$.

A semi-linear inversion (SLI) algorithm \citep{Warren03, Dye14}, which does not assume any {\it a priori} background morphology and allows multiple images to be simultaneously reconstructed using the same lens mass model, is adopted to characterise the lens. This is done making use of 7-GHz continuum and velocity-integrated CO\,(J:4$\to$3) flux maps. The total and stellar masses within the Einstein radius ($\theta_{\rm E}=2.18_{-0.27}^{+0.19}\,$kpc) are estimated to be, respectively, $\rm{M(<\theta_E)}=8.13_{-0.41}^{+0.33}\times10^{10}~$M$_\odot$ and $1.60_{-0.69}^{+1.14}\times10^{10}\,\rm{M_\sun}$, yielding a stellar mass contribution to the deflection effect of $19.7_{-8.5}^{+14.1}\%$.

The same algorithm is utilised to reconstruct the source plane at different wavelengths. The background source is magnified by $\mu\sim8-10$ (depending on wavelength) and is likely a merger event between two sources oriented respectively north-south (NS) and east-west (EW), with a projection angle between the two of $\sim80\deg$. There is also evidence for a tidal tail spanning tens of kpc, resembling the Antenn\ae\ merger (Fig.~\ref{fig:recon}).

The dynamical analysis, based on our source-plane CO\,(J:4$\to$3) cube, allows us to observe that one of the components is rotation-dominated, even though morphologically disturbed (Sec.~\ref{sec:dynamass}). The tension between dark plus baryonic mass ($22.3_{-5.8}^{+8.3}\times10^{10}\,$M$_\odot$) and the dynamical mass ($5.8\pm1.7\times10^{10}\,$M$_\odot$) estimated for the background source results from the dynamical analysis being sensitive to the NS component alone, as the EW component remains undetected in CO\,(J:4$\to$3) and 1.28-mm continuum maps. This tension was then used to estimate the dynamical mass of the EW component ($16.5_{-6.0}^{+8.5}\times10^{10}\,$M$_\odot$) and infer a merger mass-ratio of 1:2.8$_{-1.5}^{+1.8}$ (Sec.~\ref{sec:dynsedmass}).

The system as a whole has a stellar mass of $1.32_{-0.41}^{+0.63}\times10^{11}\,$M$_\odot\,$, it is actively forming stars (SFR of 394$_{-88}^{+91}\,$M$_\odot\,$yr$^{-1}$ and specific SFR of 3.0$_{-1.3}^{+2.1}\,$Gyr$^{-1}$, Sec.~\ref{sec:fbsed}), and has a significant gas reservoir in its ISM (${\rm4.6\pm1.7\times10^{10}\,M_\sun}$ comprising $\sim$25\% of the baryonic mass, Sec.~\ref{sec:molecmass}). This implies a depletion time due to star formation alone of $\tau_{\rm SF}=117_{-51}^{+51}\,$Myr, which is in agreement with that expected for the SMG phase \citep[$\sim$100\,Myr, e.g.][]{Tacconi06}.

The comparison between SFRs computed via FIR/millimetre and radio estimators yields no strong evidence for active galactic nucleus activity.

Thanks to a plethora of multi-wavelength datasets, it was possible to have a first glimpse of the properties of H1429$-$0028. A glimpse of time was what actually took ALMA --- still in Cycle-0 --- to provide the rich set of information at mm wavelengths, showing how efficient can be the teaming of \emph{Herschel}-ATLAS with ALMA to find and study these rare, fortuitous events, enabling the unprecedented detailed assessment of galaxy mass assembly mechanisms with cosmic time.

\begin{acknowledgements}
HM acknowledges the support by CONYCIT-ALMA through a post-doc scholarship under the project 31100008. HM acknowledges support by FCT via the post-doctoral fellowship SFRH/BPD/97986/2013 and the program PEst-OE/FIS/UI2751/2014. 
NN and RD acknowledge support from BASAL PFB-06/2007, Fondecyt 1100540 and Anillo ACT1101.
RJI, SJM and LD ackowledge support from the European Research Council in the form of Advanced Investigator grant, {\sc cosmicism}. 
YKS acknowledges support by FONDECYT Grant No. 3130470. 
KR acknowledges support from the European Research Council Starting Grant SEDmorph (P.I. V.~Wild).\\

This paper makes use of the following ALMA data: ADS/JAO.ALMA\#2011.0.00476.S. ALMA is a partnership of ESO (representing its member states), NSF (USA) and NINS (Japan), together with NRC (Canada) and NSC and ASIAA (Taiwan), in cooperation with the Republic of Chile. The Joint ALMA Observatory is operated by ESO, AUI/NRAO and NAOJ. The National Radio Astronomy Observatory is a facility of the National Science Foundation operated under cooperative agreement by Associated Universities, Inc.\\

This publication is based on data acquired with the Atacama Pathfinder Experiment (APEX). APEX is a collaboration between the Max-Planck-Institut fur Radioastronomie, the European Southern Observatory, and the Onsala Space Observatory.\\

Support for CARMA construction was derived from the Gordon and Betty Moore Foundation, the Kenneth T. and Eileen L. Norris Foundation, the James S. McDonnell Foundation, the Associates of the California Institute of Technology, the University of Chicago, the states of California, Illinois, and Maryland, and the National Science Foundation. Ongoing CARMA development and operations are supported by the National Science Foundation under a cooperative agreement, and by the CARMA partner universities.\\

Based in part on observations obtained at the Gemini Observatory [include additional acknowledgement here, see section 1.2], which is operated by the Association of Universities for Research in Astronomy, Inc., under a cooperative agreement with the NSF on behalf of the Gemini partnership: the National Science Foundation (United States), the National Research Council (Canada), CONICYT (Chile), the Australian Research Council (Australia), Ministério da Ciência, Tecnologia e Inovação (Brazil) and Ministerio de Ciencia, Tecnología e Innovación Productiva (Argentina).\\

The Herschel-ATLAS is a project with {\it Herschel}, an ESA space observatory with science instruments provided by European-led Principal Investigator consortia and with important participation from NASA. The H-ATLAS website is http://www.h-atlas.org/.\\

Based in part on observations carried out with the IRAM 30m Telescope. IRAM is supported by INSU/CNRS (France), MPG (Germany) and IGN (Spain).\\

Some of the data presented herein were obtained at the W.M. Keck Observatory, which is operated as a scientific partnership among the California Institute of Technology, the University of California and the National Aeronautics and Space Administration. The Observatory was made possible by the generous financial support of the W.M. Keck Foundation.\\

The authors wish to recognize and acknowledge the very significant cultural role and reverence that the summit of Mauna Kea has always had within the indigenous Hawaiian community. We are most fortunate to have the opportunity to conduct observations from this mountain.\\

Funding for SDSS-III has been provided by the Alfred P. Sloan Foundation, the Participating Institutions, the National Science Foundation, and the U.S. Department of Energy Office of Science. The SDSS-III web site is http://www.sdss3.org/.\\

SDSS-III is managed by the Astrophysical Research Consortium for the Participating Institutions of the SDSS-III Collaboration including the University of Arizona, the Brazilian Participation Group, Brookhaven National Laboratory, University of Cambridge, Carnegie Mellon University, University of Florida, the French Participation Group, the German Participation Group, Harvard University, the Instituto de Astrofisica de Canarias, the Michigan State/Notre Dame/JINA Participation Group, Johns Hopkins University, Lawrence Berkeley National Laboratory, Max Planck Institute for Astrophysics, Max Planck Institute for Extraterrestrial Physics, New Mexico State University, New York University, Ohio State University, Pennsylvania State University, University of Portsmouth, Princeton University, the Spanish Participation Group, University of Tokyo, University of Utah, Vanderbilt University, University of Virginia, University of Washington, and Yale University.\\

This work is based in part on observations made with the Spitzer Space Telescope, which is operated by the Jet Propulsion Laboratory, California Institute of Technology under a contract with NASA.\\

This publication makes use of data products from the Wide-field Infrared Survey Explorer, which is a joint project of the University of California, Los Angeles, and the Jet Propulsion Laboratory/California Institute of Technology, funded by the National Aeronautics and Space Administration.\\

The authors thank the ALMA contact scientist Adam Leroy for the help throughout scheduling block preparation and quality assurance, and the help provided by Alexander J. Conley and Elisabete da Cunha handling, respectively, the {\sc emcee} and {\sc magphys} codes.

\end{acknowledgements}

\end{document}